\documentclass{pasj01}
\usepackage{graphicx}	
\usepackage[utf8]{inputenc}
\Received{$\langle$reception date$\rangle$}
\Accepted{$\langle$acception date$\rangle$}
\Published{$\langle$publication date$\rangle$}

\newcommand{\teff}{$T_{\rm eff}$}
\newcommand{\logg}{$\log g$}
\newcommand{\vsini}{$v \sin i$}
\newcommand{\kms}{km\,s$^{-1}$}
\newcommand{\ds}{$\delta$ Scuti}

\newcommand{\gd}{$\gamma$\,Doradus}

\begin{document}

\title{Spectroscopic and Photometric Investigation of Some Potentially Chemically Peculiar $\delta$ Scuti Stars}
\author{Filiz \textsc{Kahraman Al\.{ı}\c{c}avu\c{s}}\altaffilmark{1,2}\thanks{E-mail: filizkahraman01@gmail.com}, Fahri \textsc{Al\.{ı}\c{c}avu\c{s}}\altaffilmark{1,2}, Burcu \textsc{\"{O}zkarde\c{s}}\altaffilmark{2,3}, Eda \textsc{\c{C}el\.{i}k}\altaffilmark{4}}
\altaffiltext{1}{\c{C}anakkale Onsekiz Mart University, Faculty of Science, Physics Department, TR-17100, \c{C}anakkale, T\"{u}rkiye}
\altaffiltext{2}{\c{C}anakkale Onsekiz Mart University, Astrophysics Research Center and Ulup{\i}nar Observatory, TR-17100, \c{C}anakkale, T\"{u}rkiye}
\altaffiltext{3}{\c{C}anakkale Onsekiz Mart University, Faculty of Science, Space Sciences and Technologies Department, TR-17100, \c{C}anakkale, T\"{u}rkiye}
\altaffiltext{4}{\c{C}anakkale Onsekiz Mart University, School of Graduate Studies, Physics Department, TR-17100, \c{C}anakkale, T\"{u}rkiye}


\maketitle

\begin{abstract}

Investigating chemically peculiar pulsating stars is crucial for understanding the pulsation driving mechanism in detail. To reveal the true peculiarity properties of stars detailed spectroscopic analysis is essential. Therefore, in this study, we focused on \ds\, stars previously identified as chemically peculiar but needed comprehensive updated spectroscopic analysis to uncover the chemical abundance structure of them. We selected ten targets which have public high-resolution spectroscopic and photometric data. Performing spectral analyses, we determined the spectral classification, atmospheric parameters, and detailed chemical abundance distributions of the selected stars. The pulsation properties were also analyzed using TESS data and pulsation modes for the highest amplitude pulsation frequencies were derived. We estimated the masses and ages of the targets using the evolutionary tracks and isochrones. As a result of the study, we show that only three targets exhibit chemical peculiarity: AU\,Scl and FG\,Eri as metallic A (Am) stars, and HZ\,Vel as a $\lambda$\,Bootis. However, others were found to be chemically normal stars. This study show us the importance of chemical abundance analysis in the classification of chemical peculiar stars.

\end{abstract}
\section{Introduction}

Stars that exhibit chemical abundance patterns different from the solar abundance are known as chemically peculiar stars. There are various types of chemically peculiar stars, ranging from spectral types B to F, and they can primarily be classified on the basis of their magnetic field properties. According to \citet{1974ARA&A..12..257P}, chemically peculiar stars can be categorized into four main classes: metallic A (Am), Ap, HgMn, and helium-weak stars. Additionally, there are other subclasses of chemically peculiar stars, such as $\lambda$ Bootis stars \citep{1998CoSka..27..413F} and helium-strong objects \citep{1977A&AS...30...11P}. The Am, Ap and $\lambda$ Bootis stars share a similar temperature type, approximately between A and F \citep{2024IAUS..376...91P}, but they exhibit distinct chemical compositions. Am stars show an overabundance of iron-peak elements while having weak Ca and Sc lines \citep{1970PASP...82..781C}. The chemical peculiarities in Am stars thought to be mainly caused by diffusion of elements and the loss of the outer convection zone due to helium settling \citep{2005A&A...443..627T, 2010A&A...523A..40A}. Ap stars exhibit an overabundance of certain elements such as Eu, Sr, and Cr and they are known by their strong magnetic fields \citep{2009ssc..book.....G, 2015MNRAS.454.3143A}. The chemical peculiarity in this systems is caused by the magnetic diffusion processes which tend to raise certain chemical elements to accumulate near the magnetic poles \citep{2015MNRAS.454.3143A}. In contrast of Am and Ap stars, $\lambda$ Bootis stars display a metal-deficient atmospheric structure \citep{1998CoSka..27..413F}.  In the case of $\lambda$\,Bootis stars, their peculiar chemical patterns are thought to result from the accretion of material from the interstellar medium \citep{2002MNRAS.335L..45K}.

\begin{table*} 
    \centering 
      \caption{Information about the targets. The spectral types given in this table were taken from literature studies as cited in the last column.}\label{table1}
    \begin{tabular}{rllllll}
    \hline
HD     &Star  & RA   & DEC  & V     & Spectral &References  \\
Number &Name  & (deg)& (deg)& (mag) & type     &\\     
        \hline
1097 & AU\,Scl  & 3.78 & -29.01  & 9.05 & A3/5mF0-F5 & \citet{1982mcts.book.....H} \\
11956 & FG\,Eri  & 28.95 & -55.07  & 6.71 & kF0hA5mF0V& \citet{2001A&A...373..625P} \\
23194 & V1187 Tau & 56.00  & 24.56 & 8.07 & A4-A7m & \citet{2009A&A...498..961R}\\
75654 & HZ\,Vel    & 132.47 & -39.14  & 6.37 & hF0mA5V& \citet{2001A&A...373..633P}\\
95321 & V527\,Car & 164.84  & -58.12 & 9.10 & A3mA7-A9 & \citet{1975mcts.book.....H} \\
104513 & DP\,UMa & 180.53  & 43.05 & 5.21 & F0Vam & \citet{2000A&AS..144..469R}\\
107513 & KU\,Com & 185.36  & 24.99 & 7.38 & kA7hF0mF0IV & \citet{2018MNRAS.480.2953G}\\
125081 & MX\,Vir & 214.42  & -21.83 & 7.35 & kF2hF5mF5II & \citet{2009A&A...498..961R} \\
 &  & &  &  &F3 Sr Cr Eu & \\
143232 & IO\,Lup & 240.19  & -39.09 & 6.66 & A5mA5-F2 & \citet{2000A&AS..144..469R}\\
213204 & UV\,PsA & 337.55  & -30.44 & 8.39 & F1m& \citet{2011A&A...535A...3S}\\
        \hline
    \end{tabular}
\end{table*}

Some chemically peculiar stars were found to display oscillations \citep{2011A&A...535A...3S, 2017MNRAS.466..546M}. One such class is the \ds\, variables, which are A-F type stars pulsating in the frequency range of ~5-80\,d$^{-1}$ with pressure and mixed modes \citep{2000ASPC..210....3B, 2010aste.book.....A, 2013AJ....145..132C}. These variables lie within their own instability strip, overlapping with the instability domain of \gd\, stars. \gd\, stars, typically F-type, oscillate at lower frequencies (below 5\,d$^{-1}$) in gravity modes \citep{1999PASP..111..840K}. Within the overlapping region of the instability strips of these pulsating stars, hybrid pulsators, which simultaneously exhibit the pulsation characteristics of both types of variables, are located \citep{2002MNRAS.333..251H, 2011A&A...534A.125U}.

Pulsations in Am stars were initially unexpected due to atomic diffusion and gravitational element settling \citep{1976ApJS...32..651K, 2000A&A...360..603T}. However, it is now well established that many \ds\, and \gd\, stars exhibit chemical peculiarities \citep{2017MNRAS.465.2662S, 2017MNRAS.470.2870N, 2019MNRAS.490.4040A}. Investigating these peculiar pulsating stars is crucial to understanding how chemical anomalies influence oscillations. \citet{2017MNRAS.465.2662S} studied a large sample of $\delta$ Scuti Am stars using low-resolution LAMOST spectra and ground-based photometric data from the WASP (the Wide Angle Search for Planets). They found that $\delta$ Scuti Am stars are predominantly located near the cool edge of the $\delta$ Scuti instability strip. The effect of metallicity on pulsation was also investigated in this study; however, no significant correlation between oscillation properties and metallicity was found. Moreover, $\delta$ Scuti-type pulsations in Am stars could not be fully explained by the conventional Kappa mechanism alone. In the study by \citet{2014ApJ...796..118A}, the pulsations in these stars were explained for the first time by turbulent pressure, a result later supported by \citet{2017MNRAS.465.2662S}. More recently, \citet{2020MNRAS.498.4272M} and \citet{2024A&A...690A.104D} provided further evidence that pulsations in chemically peculiar stars are more common than previously believed. \citet{2020MNRAS.498.4272M} demonstrated that \ds\, type pulsations can exist with rapid oscillations in magnetic Ap stars, even under significant helium depletion and moderate magnetic field strengths. Additionally, \citet{2024A&A...690A.104D} analyzed a large sample of metallic A and F stars using the Transiting Exoplanet Survey Satellite (TESS) and Gaia data and found that a substantial fraction of them exhibit pressure mode pulsations. Their results show that many of these pulsating chemically peculiar stars are located near the main sequence and cluster toward the red edge of the classical instability strip.

Even if some stars classified as \ds\, Am in the literature \citep{2000A&AS..144..469R}, many of their classifications rely on photometric analysis and/or spectral analysis without detailed chemical abundance studies and this may cause misclassifications. These results could negatively affect on the investigation about pulsating chemically peculiar stars. Therefore, to control peculiarity classification, we selected some target stars for in-depth analysis to accurately determine their atmospheric parameters and chemical compositions.

We determined the targets based on several criteria. They were selected from the catalog of \ds\, stars \citep{2013AJ....145..132C} through a review of the literature. First, we focused on \ds\, stars that are identified as chemically peculiar based on their spectral types, and for which there is a need of detailed spectroscopic study for determinations chemical abundances of individual elements available in the literature. Additionally, we only considered stars with available public spectroscopic data. Following this process, a set of potential targets was identified, and from this group, only those with TESS data were selected to investigate pulsation properties in detail with the same source of data for all targets. Information about the targets is provided in Table\,\ref{table1} and literature information of the selected targets is given in Appendix. The paper is organized as follows. In Sect.\,2, details about photometric and spectroscopic data are given. The spectral analysis to determined the spectral classification, atmospheric parameters and also the chemical abundances of individual elements is presented in Sect.\,3. The photometric investigation is also introduced in Sect.\,4. Discussion and Conclusions are given in the last section of Sect.\,5.

\section{Photometric and spectroscopic data}

\begin{figure}
\centering
  \centering
  \includegraphics[alt={TESS pixel file for V1187 Tau}, height=5cm, width=0.5\textwidth]{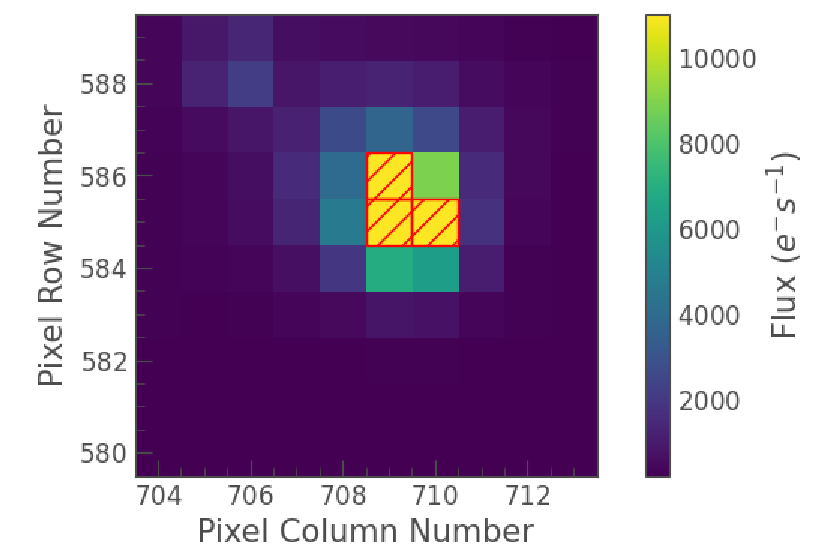}
\caption{TESS target pixel file image for V1187\,Tau. The color scale represents the flux intensity recorded in each pixel. The over-plotted red hatched region indicates the photometric aperture used to extract the light curve, centered on the main flux-contributing source.}
\label{fig:pixelfile}
\end{figure}

The TESS photometric data for the targets were used in the analysis. TESS provides at least approximately 27 days of data for each target, depending on the objects's position in the sky \citep{2015JATIS...1a4003R}. These photometric data are available in the Mikulski Archive for Space Telescopes (MAST)\footnote{https://mast.stsci.edu/}, with varying fluxes and exposure lengths depending on the specific TESS observations. The exposure times of the TESS data are 120s, 200s, 600s, and 1800s. Since our targets are \ds\, type pulsators with pulsation frequencies ranging around from 5 to 80 d$^{-1}$ \citep{2000A&AS..144..469R, 2013AJ....145..132C}, and considering the Nyquist frequency of the TESS data, we chose to use the 120s TESS data for our analysis. We selected the pre-search aperture photometry (PDCSAP) data for the analysis \citep{2015JATIS...1a4003R}. The available TESS sector data for our targets are listed in Table \ref{tab:table2}. Since TESS pixels are relatively large (21"), there is a possibility of contamination from nearby sources. To minimize this, we examined the TESS pixel-level data and ensured that the extracted flux predominantly originates from our target. Fig.\,\ref{fig:pixelfile} shows an example of a pixel file for one of our targets (V1187\,Tau), where the flux is clearly concentrated within the selected aperture and no significant contamination from nearby sources is evident.

\begin{table}
    \centering 
      \caption{Observational information. S/N is the abbreviation of signal-to-noise. The S/N value for DP UMa is given considering the combined spectrum as explained in Sect.\,2.}\label{tab:table2}
    \begin{tabular}{lllll}
    \hline
Star  & TESS    & Spectrograph & S/N\\
Name  & sector &              & ratio\\     
        \hline
AU\,Scl   &  29                & HARPS  & 120\\
FG\,Eri   &  2, 3, 29, 30, 69  & HARPS & 190\\
V1187 Tau &  42, 43, 44, 70,71 & HARPS  & 170\\
HZ\,Vel   & 8, 9, 35, 62       & HARPS & 160\\
V527\,Car &  37                & HARPS & 140\\
DP\,UMa   & 22, 49, 76         & SOPHIE & 490\\
 KU\,Com  &  22, 49            & ELODIE & 210\\
MX\,Vir   &  1                 & HARPS& 170\\
 IO\,Lup  &  65                & HARPS& 160\\
 UV\,PsA  & 1, 28              & HARPS& 150\\
        \hline
    \end{tabular}
\end{table}

Spectroscopic data for the selected targets were retrieved from public spectral databases. High-resolution spectra for the targets were found in the European Southern Observatory (ESO) Science Archive Facility (SAF)\footnote{http://archive.eso.org/cms.html}, ELODIE\footnote{http://atlas.obs-hp.fr/elodie/}, and SOPHIE\footnote{http://atlas.obs-hp.fr/sophie/} archives. The ESO SAF includes spectra taken with spectrographs attached to ESO telescopes at the La Silla Paranal Observatory. One of the ESO spectrograph is the High Accuracy Radial Velocity Planet Searcher (HARPS), an \'{e}chelle spectrograph with a resolving power of 80,000, which provides spectra within a wavelength range of 378-691\,nm \citep{2003Msngr.114...20M}. The ELODIE and SOPHIE \'{e}chelle spectrographs were mounted on telescopes at the Observatoire de Haute–Provence (OHP). These have resolutions of 42,000 and 75,000, respectively, and offer spectra in the wavelength ranges 385-680\,nm for ELODIE and 387-694\,nm for SOPHIE \citep{2004PASP..116..693M, 2008SPIE.7014E..0JP}. 

Except for DP\,UMa, all targets have only a single spectrum. When DP\,UMa was selected for this study, six spectra taken in 2020 were available in the SOPHIE archive. Therefore, we used these six spectra in our analysis for this target. After collecting the spectra of the targets they were normalized using the SUPPNET program \citep{2022A&A...659A.199R}. After normalization, the spectra of each target were examined for possible effects of a binary component (if present). In addition, a cross-correlation technique was applied using the IRAF\footnote{http://iraf.noao.edu/} FXCOR package \citep{1986SPIE..627..733T}, but no evidence of a secondary component was found in the spectra. The normalized spectra of DP\,UMa were combined to obtain a combined spectrum with higher signal-to-noise ratio. The Details about the spectroscopic data and the signal-to-noise (S/N) ratio of the spectra are given in Table\,\ref{tab:table2}.

\section{Spectral analysis}

The spectroscopic data of the targets were used to estimate their atmospheric parameters such as effective temperature (\teff), surface gravity (\logg), chemical composition, and projected rotational velocity (\vsini). With this information, we can understand the chemical nature of the targets and reclassify them based on their chemical compositions.  

Before determining the atmospheric parameters and the chemical composition, we performed spectral classification for the targets. The spectral classification provides preliminary and crucial information about the stars' chemical properties and atmospheric parameters \citep{2009ssc..book.....G, 2014dap..book.....N}. The spectral classification was used to estimate the temperature type (e.g., A0, A2) and luminosity class (e.g., V, IV) of the targets by comparing the observed spectra with standard stars, as outlined in \citet{2009ssc..book.....G}. During this process, we specifically compared the hydrogen, metal, and CaIIK lines (if available) to ensure accurate classification, particularly for chemically peculiar stars. This method is detailed in \citet{2016MNRAS.458.2307K}. The estimated spectral classification are listed in Table\,\ref{tab:tab3_pho}.

As observed from the spectral classification, some stars exhibit different temperature types depending on whether hydrogen (h), metal (m), or CaIIK (k) lines are considered. If there is a significant subclass difference between the temperature types derived from metal and Ca II K lines, and if the metal lines suggest a cooler temperature type, the star is classified as a metallic-line (Am) star. Conversely, if the temperature type derived from the metal lines indicates a hotter type, the star is categorized as a metal-weak star. According to the spectral classification presented in Table\,\ref{tab:tab3_pho}, AU\,Scl and FG\,Eri looks exhibiting Am type spectral classification, while HZ\,Vel seems a metal-weak star. Even though spectral classification provides an idea about the \teff\, values and chemical structure of stars, it should be noted that this method is subjective and involves higher uncertainties comparing to spectral analysis. Therefore, for a reliable classification of peculiarities, determining the chemical abundance pattern of the stars is the most accurate approach.

\begin{table*}
    \centering
    \caption{The $E(B-V)$ values, updated spectral classfication, and atmospheric parameters derived from photometric indices. The standard error for $E(B-V)$ is 0.002 mag.} \label{tab:tab3_pho}
    \begin{tabular}{lllllll}
    \hline
        Star    & Spectral  &$E(B-V)$  & \teff$^{UBV}$  & \teff$^{VK}$ & \teff$^{uvby\beta}$& \logg$^{uvby\beta}$  \\ 
         name    & class    & (mag)     & (K)       & (K)        & (K)                   & (cgs)          \\ \hline
        AU\,Scl & kA4hA3mF3\,IV/V & 0.054  & 6770\,$\pm$\,178 & 7410\,$\pm$\,80 & 7200\,$\pm$\,120 & 4.85\,$\pm$\,0.20 \\ 
        FG\,Eri & kA7hA7mF0\,V  & 0.010 & 8090\,$\pm$\,152 & 7700\,$\pm$\,120 & 8050\,$\pm$\,100 & 3.48\,$\pm$\,0.11\\ 
        V1187\,Tau & A5/6\,V & 0.031 & 7890\,$\pm$\,145 & 7910\,$\pm$\,120 & 8370\,$\pm$\,100 & 4.37\,$\pm$\,0.15 \\
        HZ\,Vel & hF0mA5\,V  &0.000 & 7540\,$\pm$\,160 & 7190\,$\pm$\,100 & 8700\,$\pm$\,150 & 3.70\,$\pm$\,0.15  \\
        V527\,Car & A3/4\,IV& 0.043 & 8110\,$\pm$\,290 & 7500\,$\pm$\,130 &  &   \\ 
        DP\,UMa & F0\,IV & 0.000 & 6800\,$\pm$\,155 & 7500\,$\pm$\,150 & 7400\,$\pm$\,120 & 4.10\,$\pm$\,0.12\\
        KU\,Com & F0\,V &0.000 & 7380\,$\pm$\,120 & 7230\,$\pm$\,100 & 7270\,$\pm$\,120 & 4.01\,$\pm$\,0.10  \\
        MX\,Vir & kF3hF5mF5\,IV-III &0.012 & 6575\,$\pm$\,100 & 6440\,$\pm$\,80 & 6850\,$\pm$\,80 & 3.70\,$\pm$\,0.10  \\
        IO\,Lup & F0\,IV & 0.005 & 7595\,$\pm$\,130 & 7580\,$\pm$\,100 & 7760\,$\pm$\,100 & 3.75\,$\pm$\,0.13 \\ 
       UV\,PsA & F2\,/IV &0.060 & 7160\,$\pm$\,170 & 7310\,$\pm$\,135 & 7000\,$\pm$\,90 & 3.77\,$\pm$\,0.12  \\          
         \hline
    \end{tabular}
\end{table*}

\subsection{Determination of the atmospheric parameters}

Before determining the spectroscopic atmospheric parameters, to have an initial value for the \teff\, and \logg, we first estimated the \teff\, and \logg\, parameters using photometric colors. Since these colors are influenced by interstellar reddening, E(B\,-\,V), we calculated these values using the dust map from \citet{2019ApJ...887...93G}. After determining the E(B\,-\,V) values for each target, we gathered the photometric colors from the Tycho-2 catalog \citep{2000A&A...355L..27H}, 2MASS catalog \citep{2003tmc..book.....C} and the Stroemgren-Crawford uvby$\beta$ photometry catalog \citep{2015A&A...580A..23P}. We estimated the photometric \teff\, values from each of the collected photometric colors using the studies of \citet{2000AJ....120.1072S}, \citet{2006A&A...450..735M} and \citet{1985MNRAS.217..305M}. The \logg\, values were also computed from the uvby$\beta$ photometry based on \citet{1985MNRAS.217..305M} study. The estimated photometric atmospheric parameters and E(B\,-\,V) values for each target are listed in Table\,\ref{tab:tab3_pho}.

\begin{figure*}
 \centering
   \begin{minipage}[b]{0.45\textwidth}
  \includegraphics[alt={Theoretical Hydrogen line fit for HZ Vel}, height=7cm, width=1\textwidth]{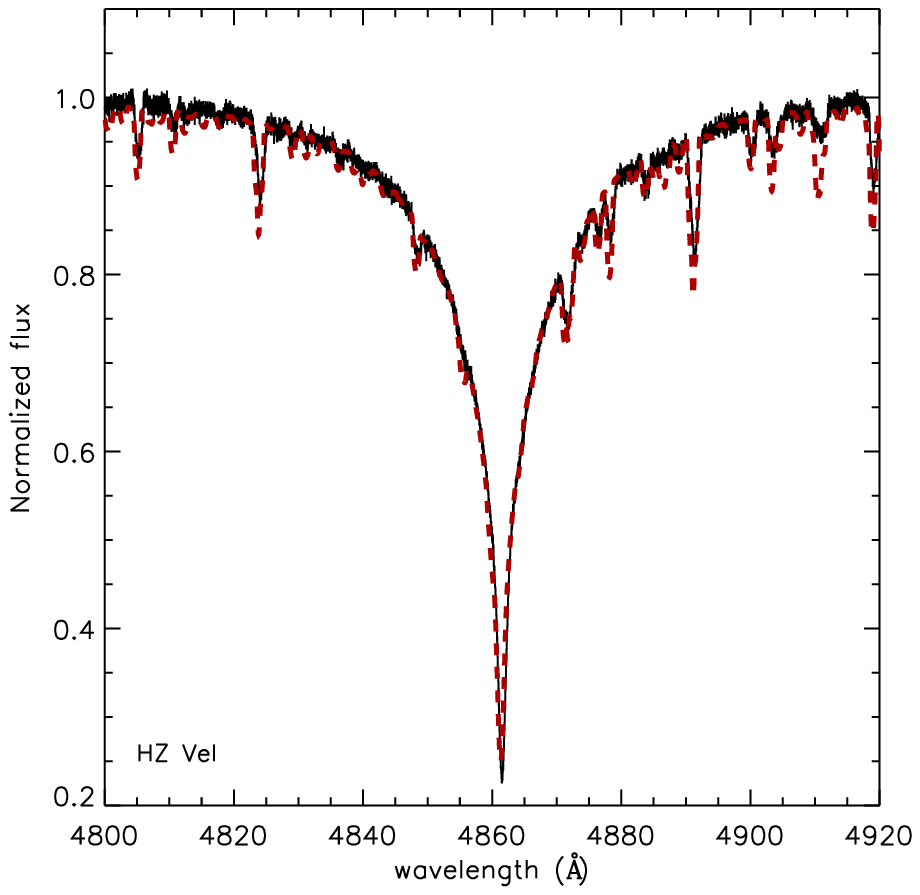}
  \end{minipage}
     \begin{minipage}[b]{0.45\textwidth}
  \includegraphics[alt={Theoretical Hydrogen line fit for IO Lup}, height=7cm, width=1\textwidth]{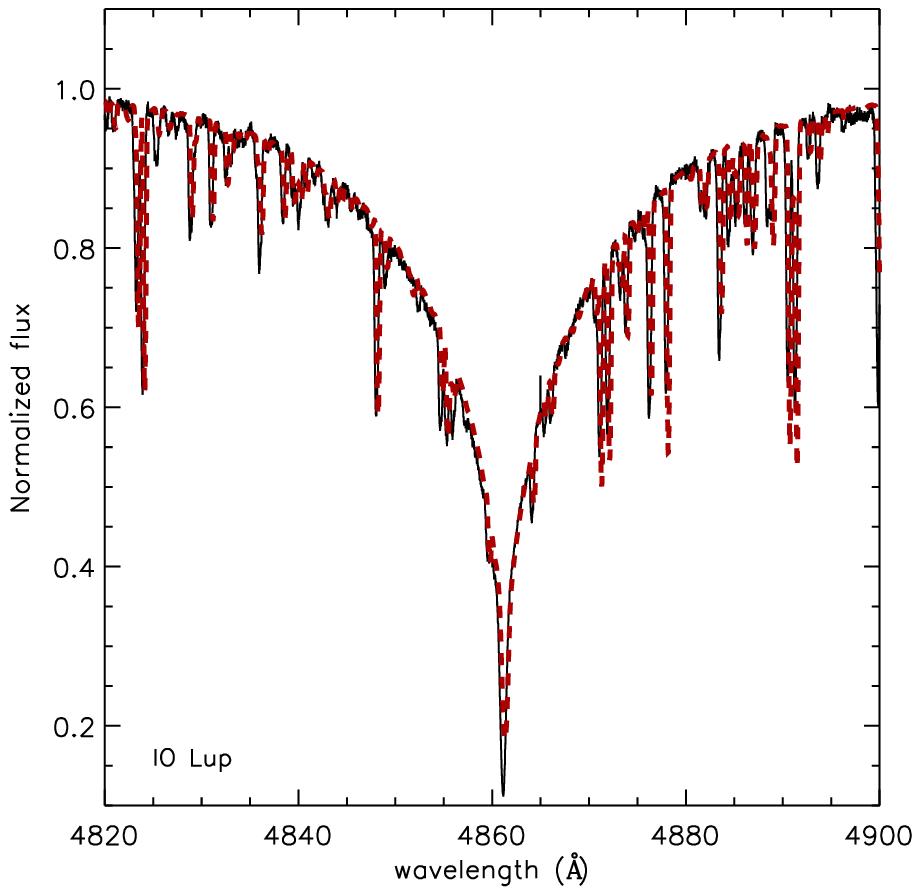}
  \end{minipage}
  \caption{The theoretical spectral (red dashed- line) fit the observed H$_{\beta}$ lines (black lines) for two of our targets.}  \label{hfits}
\end{figure*}

These photometric atmospheric parameters were then used as input for the spectroscopic analysis. In all spectroscopic analyses, we utilized the ATLAS9 model atmospheres \citep{1993KurCD..13.....K} along with the SYNTHE code \citep{1981SAOSR.391.....K} to generate synthetic spectra. Initially, hydrogen Balmer lines were used to estimate \teff. If the star was hotter than 8000\,K, the \logg\,  parameters were then determined, as the effect of \logg\, is negligible for stars cooler than 8000\,K \citep{2014dap..book.....N}. For these cooler stars we fixed the \logg\, value as 4.0. In the Balmer lines analysis, the best-fitting theoretical Balmer lines were determined using the method of \citet{2004A&A...425..641C}. In addition to atmospheric parameter determination with the Balmer lines analysis, we also utilized the Saha-Boltzmann expression. According to this expression, the abundances of an element derived from different ionizations or excitations of the element must be the same. We utilized this idea and took into account the most abundant lines of iron (Fe) due to the multitude of possible transitions between energy levels in our stars' \teff\, range. During this analysis, the spectrum synthesis method was applied and also the \vsini\, values were estimated simultaneously. Details about the analysis can be found in the study of \citet{2016MNRAS.458.2307K}. The results of the spectroscopic analysis is listed in Table\,\ref{tab:featm}, and the best fitting theoretical spectra to the Balmer lines are shown in Fig.\,\ref{hfits} for two of our targets.

\begin{table*}
    \centering
      \caption{Atmospheric parameters obtained from hydrogen balmer and iron (Fe) lines analysis. The Fe/H values are given in the last column.} \label{tab:featm}
    \begin{tabular}{lcccccc}
    \hline
Star       & \teff\ $^{Balmer}$ & \teff\ $^{Fe}$    &  \logg\ $^{Fe}$ & $\xi$           & $v\sin i$          & log $\epsilon(\mathrm{Fe})$\\ 
name       &  (K)               &   (K)             &   (cgs)         & (km/s)          & (km/s)             & \\ \hline
AU\,Scl    & 7000\,$\pm$\,200   & 7300\,$\pm$\,150  & 4.4\,$\pm$\,0.2 & 2.6\,$\pm$\,0.2 & 67\,$\pm$\,4       & 8.01\,$\pm$\,0.17\\ 
FG\,Eri    & 7700\,$\pm$\,200   & 7800\,$\pm$\,200  & 4.3\,$\pm$\,0.1 & 1.8\,$\pm$\,0.2 & 110\,$\pm$\,5      & 7.67\,$\pm$\,0.16\\
V1187\,Tau & 8000\,$\pm$\,150   & 8100\,$\pm$\,100  & 4.0\,$\pm$\,0.1 & 2.3\,$\pm$\,0.2 & 37\,$\pm$\,2       & 7.30\,$\pm$\,0.15\\ 
HZ\,Vel    & 7400\,$\pm$\,150   & 7800\,$\pm$\,100  & 4.0\,$\pm$\,0.1 & 3.1\,$\pm$\,0.2 & 45\,$\pm$\,2       & 6.95\,$\pm$\,0.12\\ 
V527\,Car  & 7300\,$\pm$\,200   & 7700\,$\pm$\,150  & 4.1\,$\pm$\,0.1 & 2.8\,$\pm$\,0.1 & 78\,$\pm$\,3       & 7.28\,$\pm$\,0.15\\
DP\,UMa    & 7000\,$\pm$\,200   & 7100\,$\pm$\,100  & 4.2\,$\pm$\,0.1 & 3.2\,$\pm$\,0.2 & 72\,$\pm$\,2       & 7.31\,$\pm$\,0.15 \\ 
KU\,Com    & 7100\,$\pm$\,200   & 7100\,$\pm$\,150  & 4.0\,$\pm$\,0.1 & 3.0\,$\pm$\,0.1 & 64\,$\pm$\,3       &7.29\,$\pm$\,0.14\\ 
MX\,Vir    & 7200\,$\pm$\,100   & 7000\,$\pm$\,100  & 4.1\,$\pm$\,0.1 & 3.2\,$\pm$\,0.2 & 15\,$\pm$\,1       & 7.43\,$\pm$\,0.12 \\ 
IO\,Lup    & 7500\,$\pm$\,200   & 7500\,$\pm$\,100  & 4.0\,$\pm$\,0.1 & 2.5\,$\pm$\,0.1 & 18\,$\pm$\,2       & 6.12\,$\pm$\,0.38\\         
UV\,PsA    & 7100\,$\pm$\,150   & 7000\,$\pm$\,100  & 4.1\,$\pm$\,0.1 & 2.3\,$\pm$\,0.2 & 74\,$\pm$\,3       & 7.51\,$\pm$\,0.16\\ \hline
    \end{tabular}
\end{table*}

\subsection{Analysis of chemical abundances} \label{sect:abun}

After determining the atmospheric parameters for each target, we performed a chemical abundance analysis using these parameters, which were derived from the analysis of Fe lines, as fixed inputs. Before the chemical abundance analysis, the lines in the spectra of the targets were identified using the Kurucz line list\footnote{http://kurucz.harvard.edu/linelists.html}. Once the lines in each target's spectrum were defined, the spectrum synthesis method was employed to determine the chemical compositions. In this method, the abundance ratios of the elements responsible for the lines are adjusted until the theoretical spectrum closely matches the observed spectral lines. The ATLAS9 model atmospheres and SYNTHE code were used in this analysis as well. 

\begin{figure*}
 \centering
 \begin{minipage}[b]{0.4\textwidth}
  \includegraphics[alt={Chemical abundance distribution for AU Scl}, height=4.0cm, width=1\textwidth]{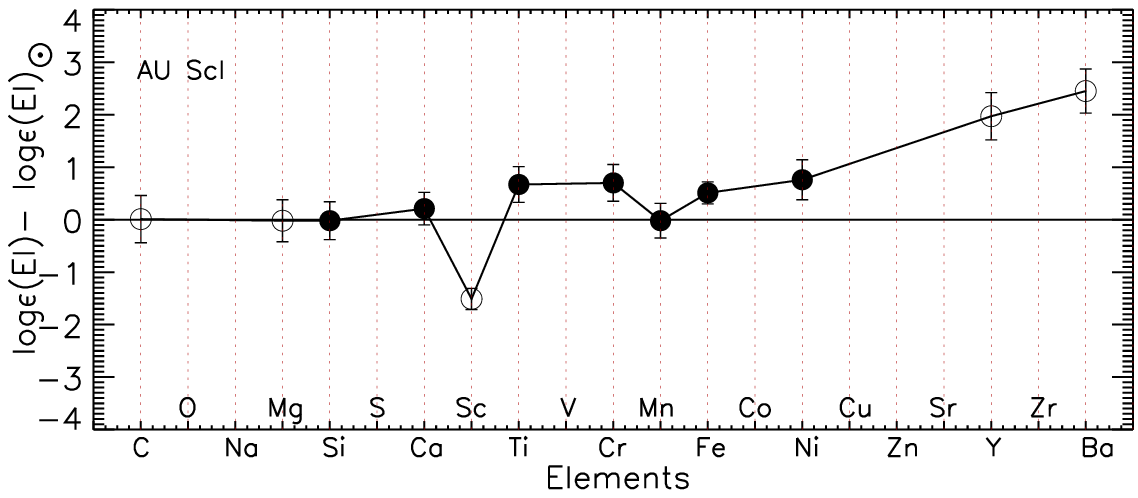}
 \end{minipage}
 \begin{minipage}[b]{0.4\textwidth}
  \includegraphics[alt={Chemical abundance distribution for FG Eri},height=4.0cm, width=1\textwidth]{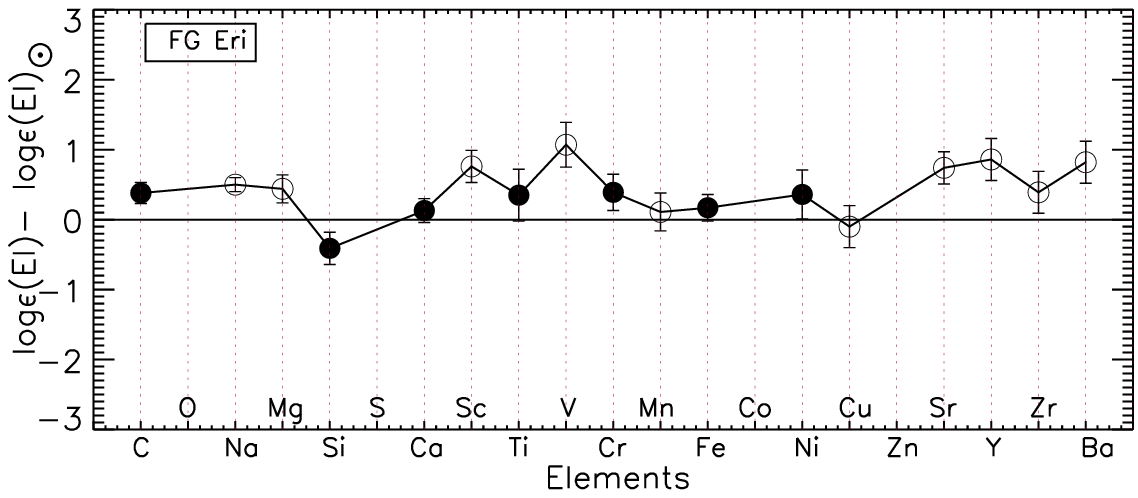}
  \end{minipage}
   \begin{minipage}[b]{0.4\textwidth}
  \includegraphics[alt={Chemical abundance distribution for V1187 Tau},height=4.0cm, width=1\textwidth]{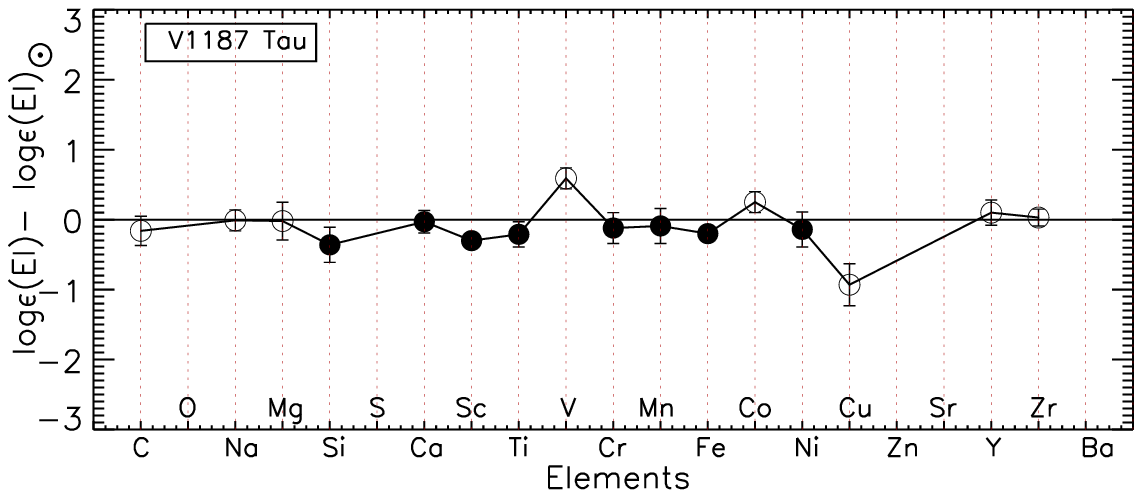}
  \end{minipage}
   \begin{minipage}[b]{0.4\textwidth}
  \includegraphics[alt={Chemical abundance distribution for HZ Vel},height=4.0cm, width=1\textwidth]{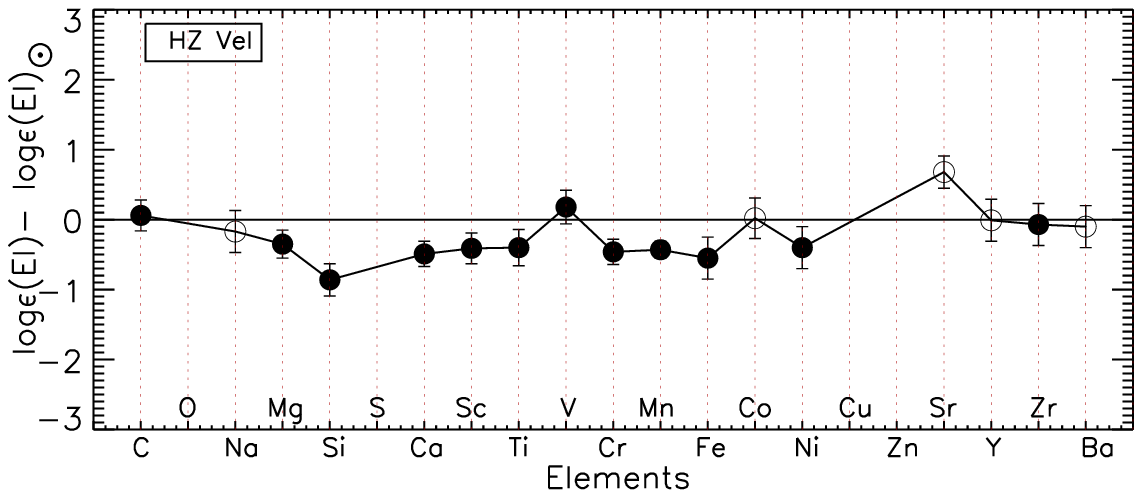}
  \end{minipage}
     \begin{minipage}[b]{0.4\textwidth}
  \includegraphics[alt={Chemical abundance distribution for V527 Car},height=4.0cm, width=1\textwidth]{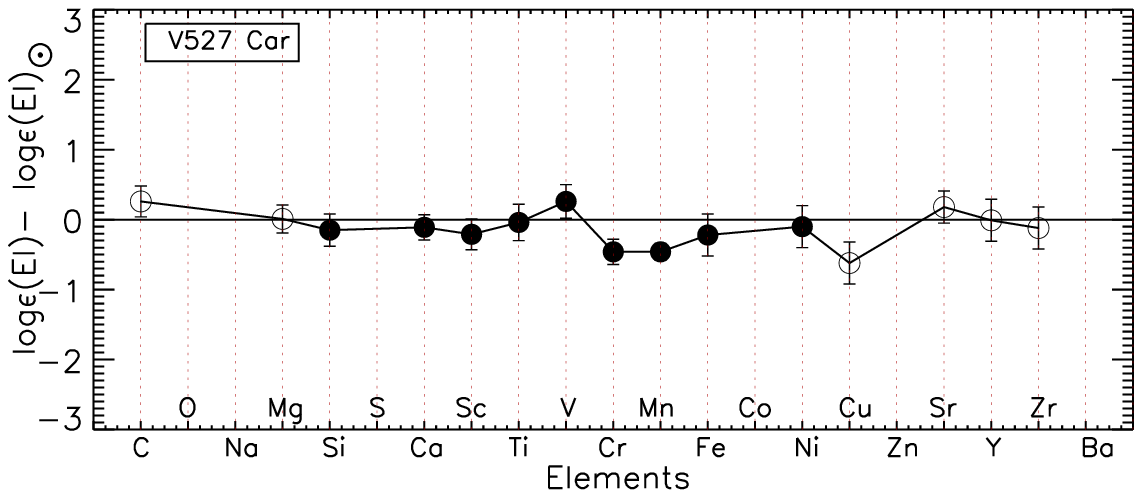}
  \end{minipage}
     \begin{minipage}[b]{0.4\textwidth}
  \includegraphics[alt={Chemical abundance distribution for DP UMa},height=4.0cm, width=1\textwidth]{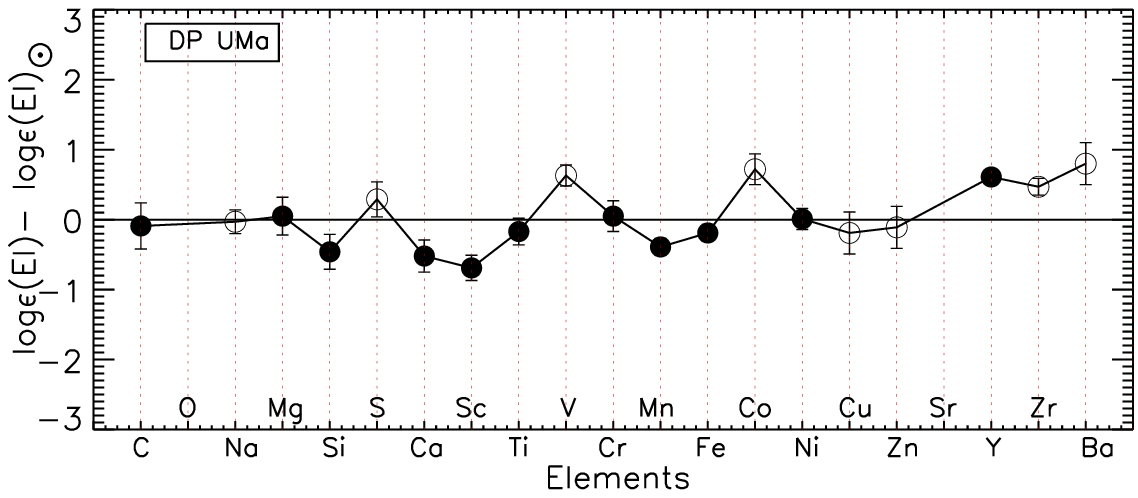}
  \end{minipage}
     \begin{minipage}[b]{0.4\textwidth}
  \includegraphics[alt={Chemical abundance distribution for KU Com},height=4.0cm, width=1\textwidth]{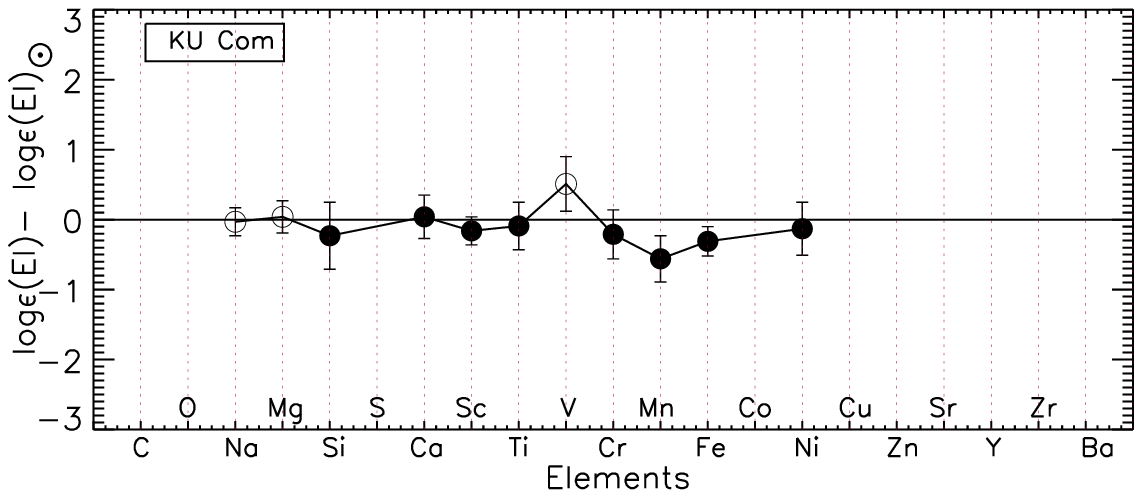}
  \end{minipage}
     \begin{minipage}[b]{0.4\textwidth}
  \includegraphics[alt={Chemical abundance distribution for MX Vir},height=4.0cm, width=1\textwidth]{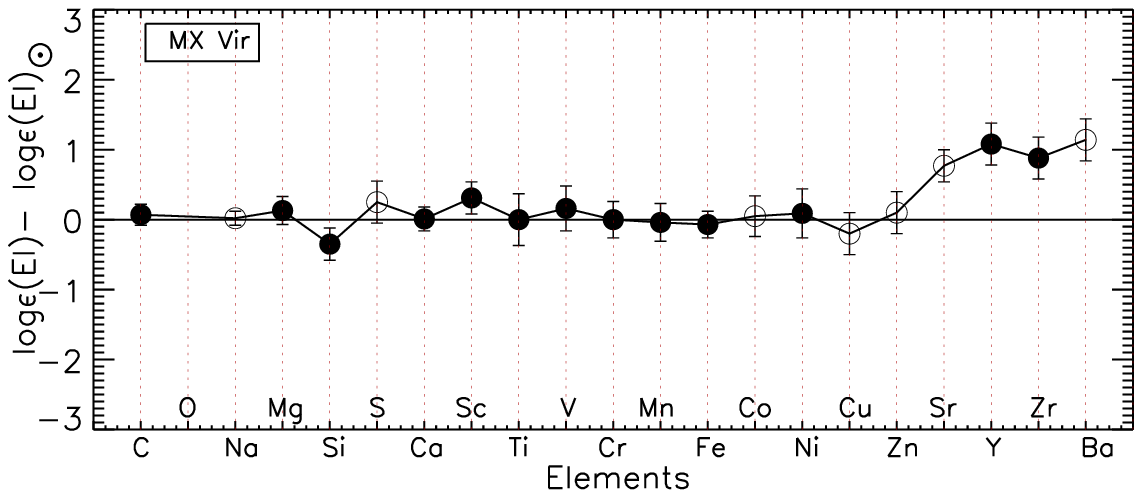}
  \end{minipage}
     \begin{minipage}[b]{0.4\textwidth}
  \includegraphics[alt={Chemical abundance distribution for IO Lup},height=4.0cm, width=1\textwidth]{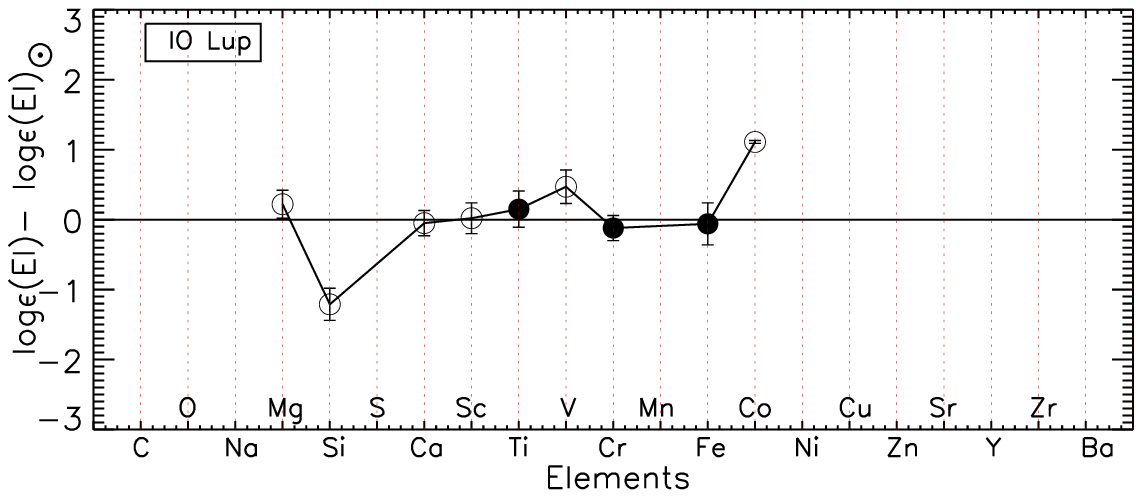}
  \end{minipage}
     \begin{minipage}[b]{0.4\textwidth}
  \includegraphics[alt={Chemical abundance distribution for UV PsA},height=4.0cm, width=1\textwidth]{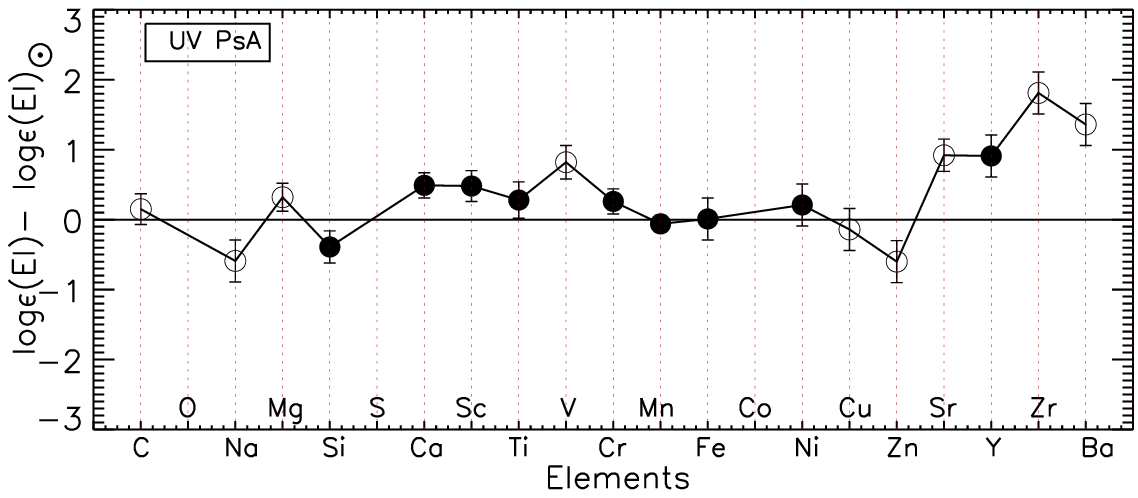}
  \end{minipage}
   \caption{Chemical abundance distribution comparing to solar abundances \citep{2009ARA&A..47..481A}. The filled and open symbols represent the number of lines used to estimate the chemical abundances of individual elements. Filled symbols indicate abundances determined from five or more lines, while open symbols correspond to abundances derived from fewer than five lines.}  \label{abundist}
\end{figure*}

As a result of the chemical abundance analysis, we determined the absolute chemical  abundance ($\log \epsilon (\mathrm{Element})$) of elements for each target. The list of absolute chemical abundances is provided in Table\,\ref{tab:table_abundance} and the [Fe/H] values for each target is listed in Table\,\ref{tab:featm}. Furthermore, the distribution of chemical abundances of the targets is shown in Fig.\,\ref{abundist}. The uncertainties in the derived absolute chemical abundances were estimated by considering the errors in the input atmospheric parameters as explained in the study of \citet{2016MNRAS.458.2307K}. For this, we systematically changed the atmospheric parameters to evaluate the impact on the derived elemental abundances in particularly in absoltute Fe abundance ($\log \epsilon (\mathrm{Fe})$). A variation of $\pm$\,100 K in \teff\, creates a change of approximately 0.07\,dex in $\log \epsilon (\mathrm{Fe})$. In comparison, $\pm$\,0.1\,dex shift in \logg\, and $\pm$\,0.1\,\kms in $\xi$ result around  about 0.01 and 0.02\,dex, respectively. We also assessed the influence of \vsini\, on abundance determinations. Our results indicate that \vsini\, introduces uncertainties ranging from $\sim$\,0.05 to 0.15\,dex. These error sources were combined in quadrature an given as uncertainties in absolute abundances (see Table\,\ref{tab:table_abundance}).

\section{Photometric analysis}

\subsection{Pulsation analysis}

The 120-s photometric TESS data of the targets were collected and they were converted into magnitude before investigating the oscillations. To convert the fluxes into the magnitudes the equation given by \citet{2022RAA....22h5003K} was used.

\begin{table}
    \centering
      \caption{The list of the highest amplitude \ds\, frequencies and their pulsation constant (Q) values.} \label{tab:modes}
    \begin{tabular}{llll}
    \hline
Star       & Pulsation & High. amp.          & Q         \\ 
name       & Class     & frequency (d$^{-1}$)& (days)    \\ \hline
AU\,Scl    & \ds\,     &16.5501              & 0.033 (6) \\ 
FG\,Eri    & Hybrid    &6.2921               & 0.060 (11)\\
V1187\,Tau & \ds\,     & 53.1771             & 0.009 (2) \\ 
HZ\,Vel    & Hybrid    & 14.3398             & 0.031 (6) \\ 
V527\,Car  & Hybrid    & 9.3603              & 0.033 (3) \\
DP\,UMa    & Hybrid    & 18.0751             & 0.032 (6) \\ 
KU\,Com    & Hybrid    &19.9108              & 0.023 (4) \\ 
MX\,Vir    & \ds\,     &6.4950               & 0.057 (10) \\ 
IO\,Lup    & \ds\,     &15.5946              & 0.020 (4) \\         
UV\,PsA    & \ds\,     & 9.1511              & 0.043 (8) \\ \hline
    \end{tabular}
\end{table}

The oscillation properties of the targets were examined using the \textsc{Period04} software \citep{2005CoAst.146...53L}, which applies a discrete Fourier transform and successive prewhitening. Given that the targets were previously identified as $\delta$\,Scuti variables (see Appendix), the frequency search was conducted in the range of 0–100\,d$^{-1}$ with a significance threshold set at 4.5$\sigma$ \citep{2021AcA....71..113B}. All available TESS sectors for each star were analyzed simultaneously. 

\begin{figure*}
\centering
  \centering
  \includegraphics[alt={Amplitude spectra for all targets}, height=20cm, width=0.9\textwidth]{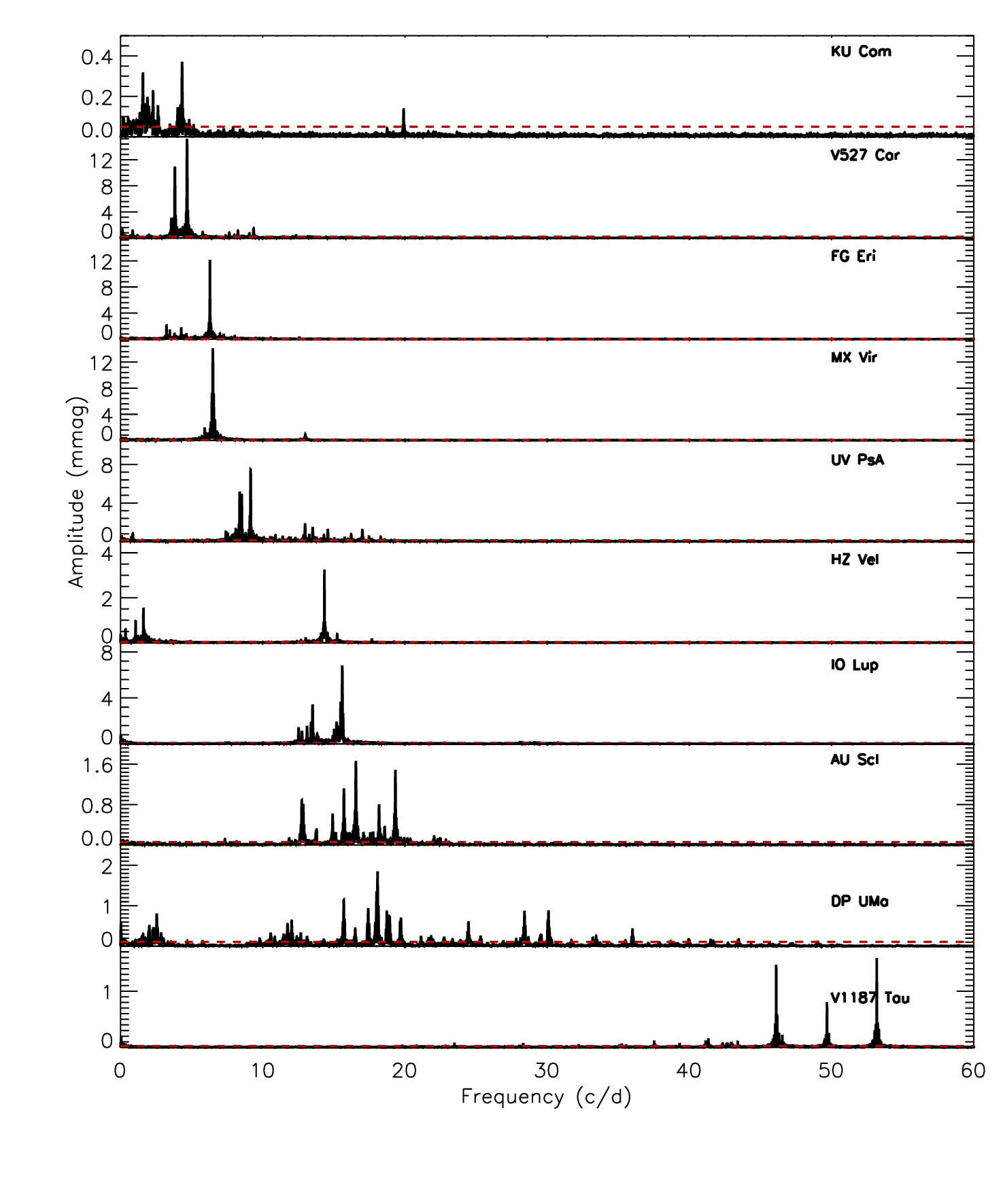}
\caption{Amplitude spectra of the targets. Dashed vertical lines represent the 4.5$\sigma$ level. }
\label{fig:puls_fre}
\end{figure*}

The significant oscillation frequencies ($f$) and their corresponding amplitudes are provided in Table\,\ref{tab1:puls_freq}, while the amplitude spectra of all targets are displayed in Fig.\,\ref{fig:puls_fre}. According to the extracted frequencies and their amplitudes, all stars display variability consistent with $\delta$\,Scuti-type pulsations. However, several stars exhibit additional low-frequency variability. Among them, AU\,Scl, MX\,Vir, IO\,Lup, UV\,PsA, and V1187\,Tau were confirmed as $\delta$\,Scuti variables. Although a few low-frequency peaks were detected in V1187\,Tau, they are likely due to small flux level shifts between sectors. Conversely, stars such as FG\,Eri, HZ\,Vel, V527\,Car, DP\,UMa, and KU\,Com show both low- and high-frequency pulsations, suggesting hybrid behavior between \ds\, and $\gamma$\,Doradus types. The pulsation classifications are listed in Table\,\ref{tab:modes}.

In addition, we computed the pulsation constant ($Q$) for the dominant frequency of each target using the classical equation of \citet{1990DSSN....2...13B}, together with the stellar parameters (e.g., $T_{\rm eff}$, $\log g$, $M_{bol}$) derived in the previous sections. The derived $Q$ values are also presented in Table\,\ref{tab:modes}. These values provide useful constraints on the nature of the dominant oscillations (e.g., low-overtone radial or non-radial modes), although definitive mode identification (i.e., determination of radial order $n$, spherical degree $l$, and azimuthal order $m$) is not attempted in this study due to the lack of spectroscopic mode diagnostics or rotational splitting analysis. Therefore, we refrain from assigning specific mode identifications and instead provide only general pulsational classifications based on the observed frequency regime and $Q$ values.

\begin{table*}
    \centering
      \caption{Estimated fundamental stellar parameters for the targets. * shows the age values taken from \citet{2021MNRAS.504..356D} and fixed during the examinations.} \label{tab:fund}
    \begin{tabular}{lccccc}
    \hline
        Star & M$_{V}$  & M$_{bol}$    &  log(L/L$_{\odot}$)     & $M$             & Age\\ 
        name  &  (mag)  &   (mag)      &                         &  (M$_{\odot}$)   & (Gyr)\\ \hline
AU\,Scl & 1.326\,$\pm$\,0.032 & 1.357\,$\pm$\,0.034 & 1.353\,$\pm$\,0.055  & 1.96\,$\pm$\,0.10 & 1.00\,$\pm$\,0.20 \\ 
FG\,Eri & -0.135\,$\pm$\,0.024 & -0.103\,$\pm$\,0.026 & 1.937\,$\pm$\,0.047& 2.48\,$\pm$\,0.12 &0.60\,$\pm$\,0.12\\
V1187\,Tau & 1.800\,$\pm$\,0.022 & 1.870\,$\pm$\,0.052 & 1.148\,$\pm$\,0.047 & 1.75\,$\pm$\,0.09 & 0.13\,$\pm$\,0.03* \\ 
HZ\,Vel & 2.087\,$\pm$\,0.023 & 2.121\,$\pm$\,0.025 & 1.047\,$\pm$\,0.046  & 1.69\,$\pm$\,0.08& 0.10\,$\pm$\,0.02\\ 
V527\,Car & 0.322\,$\pm$\,0.027 & 0.384\,$\pm$\,0.029 & 1.743\,$\pm$\,0.051 &2.23\,$\pm$\,0.11& 0.79\,$\pm$\,0.14 \\
DP\,UMa & 2.595\,$\pm$\,0.029 & 2.627\,$\pm$\,0.031 & 0.845\,$\pm$\,0.052 & 1.51\,$\pm$\,0.08 & 1.27\,$\pm$\,0.23\\ 
KU\,Com & 2.694\,$\pm$\,0.024 & 2.774\,$\pm$\,0.052 & 0.786\,$\pm$\,0.057 & 1.50\,$\pm$\,0.07& 0.63\,$\pm$\,0.19*\\ 
MX\,Vir & 1.183\,$\pm$\,0.014 & 1.263\,$\pm$\,0.043 & 1.391\,$\pm$\,0.064 &1.78\,$\pm$\,0.09 &1.47\,$\pm$\,0.27 \\ 
IO\,Lup & 0.820\,$\pm$\,0.024 & 0.854\,$\pm$\,0.026 & 1.554\,$\pm$\,0.047  & 2.08\,$\pm$\,0.10&0.97\,$\pm$\,0.19\\ 
UV\,PsA & 1.539\,$\pm$\,0.028 & 1.572\,$\pm$\,0.030 & 1.267\,$\pm$\,0.051 & 1.73\,$\pm$\,0.09&1.60\,$\pm$\,0.28\\ 
        \hline
    \end{tabular}
\end{table*}

\section{Estimating fundamental stellar parameters}

We estimated fundamental stellar parameters and the ages of the stars using evolutionary tracks and isochrones. Prior to these estimations, we also calculated other key parameters, such as luminosity ($L$), absolute magnitude (M$_{V}$), and bolometric magnitude (M$_{bol}$), using Gaia DR3 distances \citep{2023A&A...674A...1G} and bolometric corrections \citep{1996ApJ...469..355F} based on the derived \teff\, and E(B-V) values. For these calculations, we followed the method described in \citet{2021PASP..133h4201P}. The uncertainties of the calculated parameters were estimated by propagating the errors of the input parameters. These parameters and their uncertainties are listed in Table\,\ref{tab:fund}. It should be noted that, beyond these uncertainties, the possible presence of unresolved binary components may introduce additional errors in the estimated parameters of $L$, M$_{V}$ and M$_{bol}$.

For the estimation of mass ($M$) and age, we used evolutionary tracks and isochrones from MESA and Stellar Tracks (MIST) \citep{2016ApJS..222....8D, 2016ApJ...823..102C}. Details about the adopted physics used to generate the MIST evolutionary tracks and isochrones can be found in Table\,1 of \citet{2016ApJ...823..102C}. The evolutionary tracks adopted in this work assume an initial rotational velocity of $v/v_{\rm crit} = 0.4$, implying that the stars initially rotate at 40\% of their critical breakup speed. This parameterization allows for moderate rotational mixing, which can influence surface abundances and stellar lifetimes, particularly in intermediate and high mass stars. For each target, multiple evolutionary tracks and isochrones were generated, taking into account the calculated [Fe/H] values. The tracks and isochrones were computed with steps of 0.01 $M_\odot$ in $M$ and the same steps in age. Using these, we estimated the $M$ and age values by matching the observed \teff\ and $L$ values on the log\,\teff\,$-$\,log$L$ diagram. Two of our targets, V1187\,Tau and KU\,Com, are members of cluster Melotte 22 and 111, respectively. For $M$ value estimation of these systems we adopted the cluster ages from \citet{2021MNRAS.504..356D} and fixed during the examinations. The estimated masses and ages are listed in Table\,\ref{tab:fund}. The locations of the targets on the log\,\teff\,$-$\,log$L$ diagram, within the \ds\, instability strip and along the isochrones, are shown in Fig.\,\ref{evo_isc}. The uncertainties in the estimated mass and age values were derived by considering the propagated errors in the stellar parameters \teff, $\log L$, and [Fe/H], which directly affect the stars' positions on the H–R diagram and thus the resulting evolutionary track fitting. In addition to these statistical sources of uncertainty, we also included systematic lower limits following the findings of \citet{2022ApJ...927...31T}, who demonstrated that even under optimal conditions, isochrone-based determinations are subject to minimum uncertainties of $\sim$5\% in mass and $\sim$20\% in age. These systematic floors were incorporated into the final uncertainties to provide a more realistic error bars.


In addition to fundamental stellar parameters estimation we also controlled the position of these pulsating components on the \ds\, instability strip as can be seen in  Fig.\,\ref{evo_isc}. All stars in our sample were found to lie inside or close to the red edge of the \ds\, instability strip, within the uncertainties. 

\setcounter{figure}{4}
\begin{figure*}
 \centering
 \begin{minipage}[b]{0.4\textwidth}
  \includegraphics[alt={HR diagram for AU Scl},height=4.7cm, width=1\textwidth]{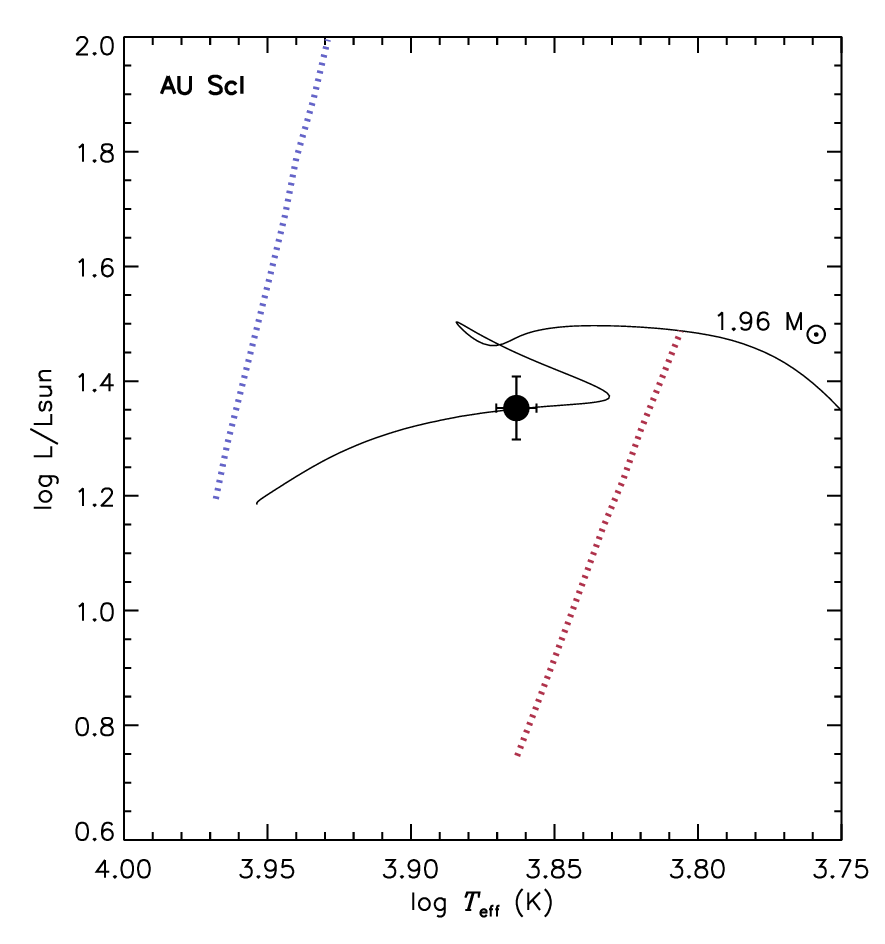}
 \end{minipage}
 \begin{minipage}[b]{0.4\textwidth}
  \includegraphics[alt={Isochrone for AU Scl},height=4.7cm, width=1\textwidth]{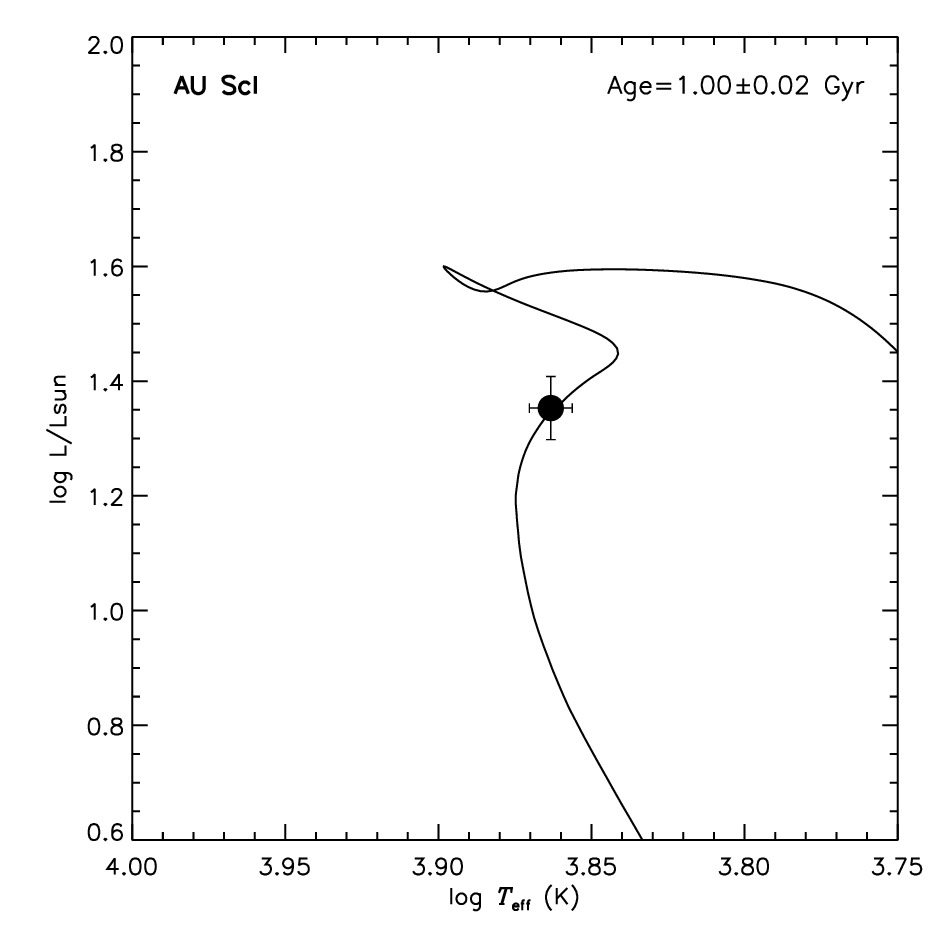}
  \end{minipage}
   \begin{minipage}[b]{0.4\textwidth}
  \includegraphics[alt={HR diagram for FG Eri},height=4.7cm, width=1\textwidth]{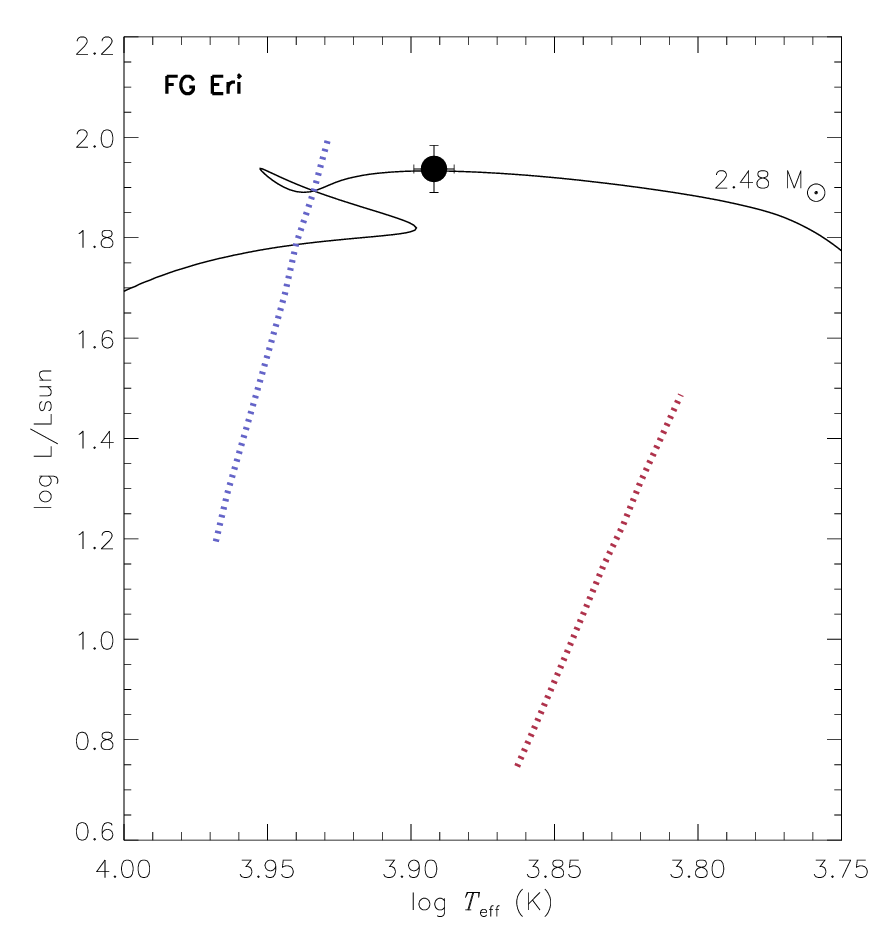}
  \end{minipage}
   \begin{minipage}[b]{0.4\textwidth}
  \includegraphics[alt={Isochrone for FG Eri},height=4.7cm, width=1\textwidth]{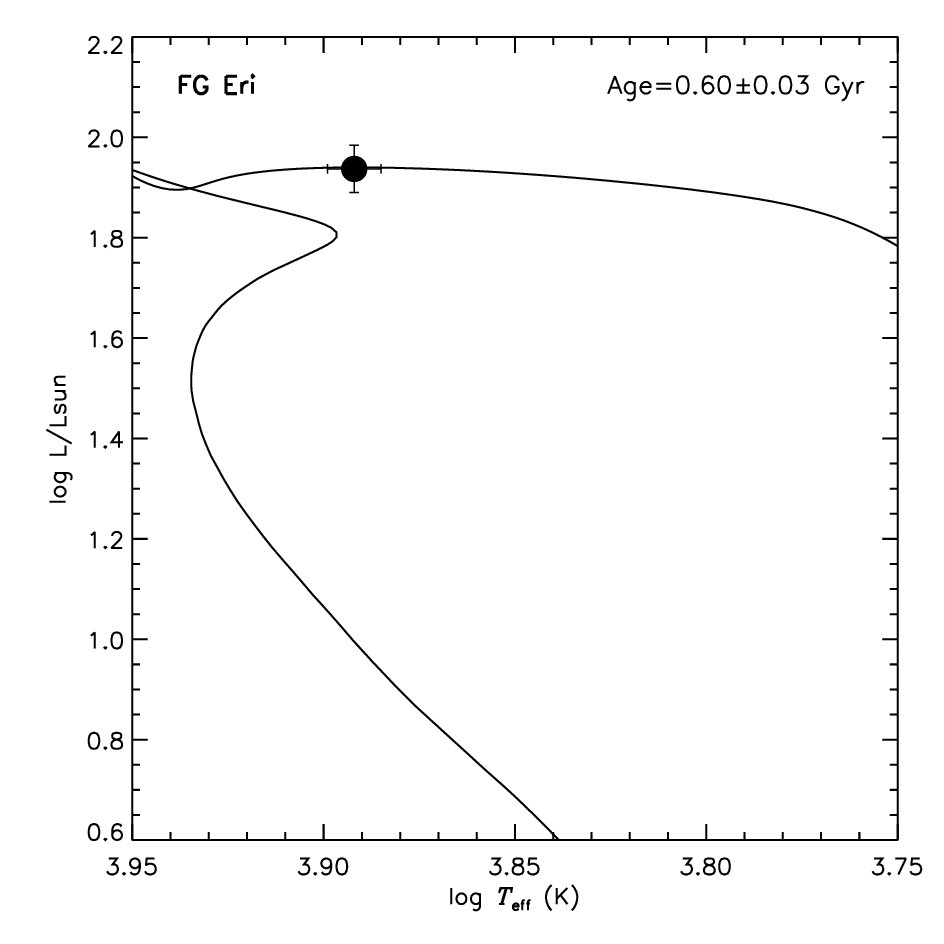}
  \end{minipage}
     \begin{minipage}[b]{0.4\textwidth}
  \includegraphics[alt={HR diagram for V1187 Tau},height=4.7cm, width=1\textwidth]{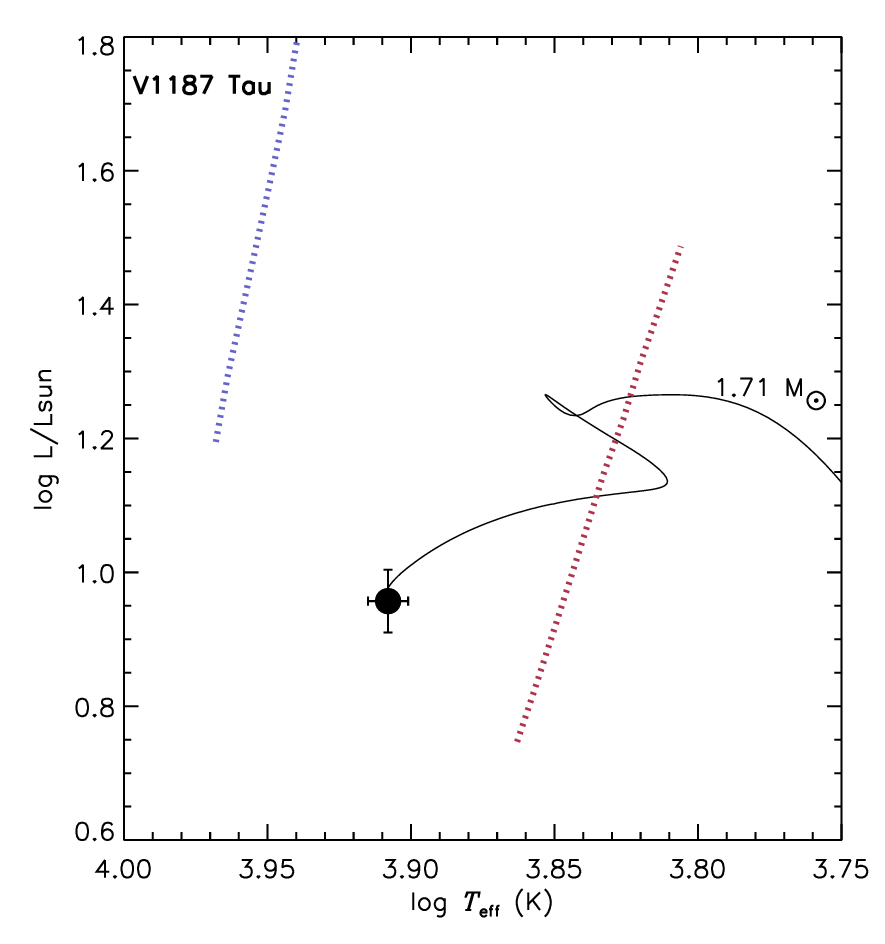}
  \end{minipage}
     \begin{minipage}[b]{0.4\textwidth}
  \includegraphics[alt={Isochrone for V1187 Tau},height=4.7cm, width=1\textwidth]{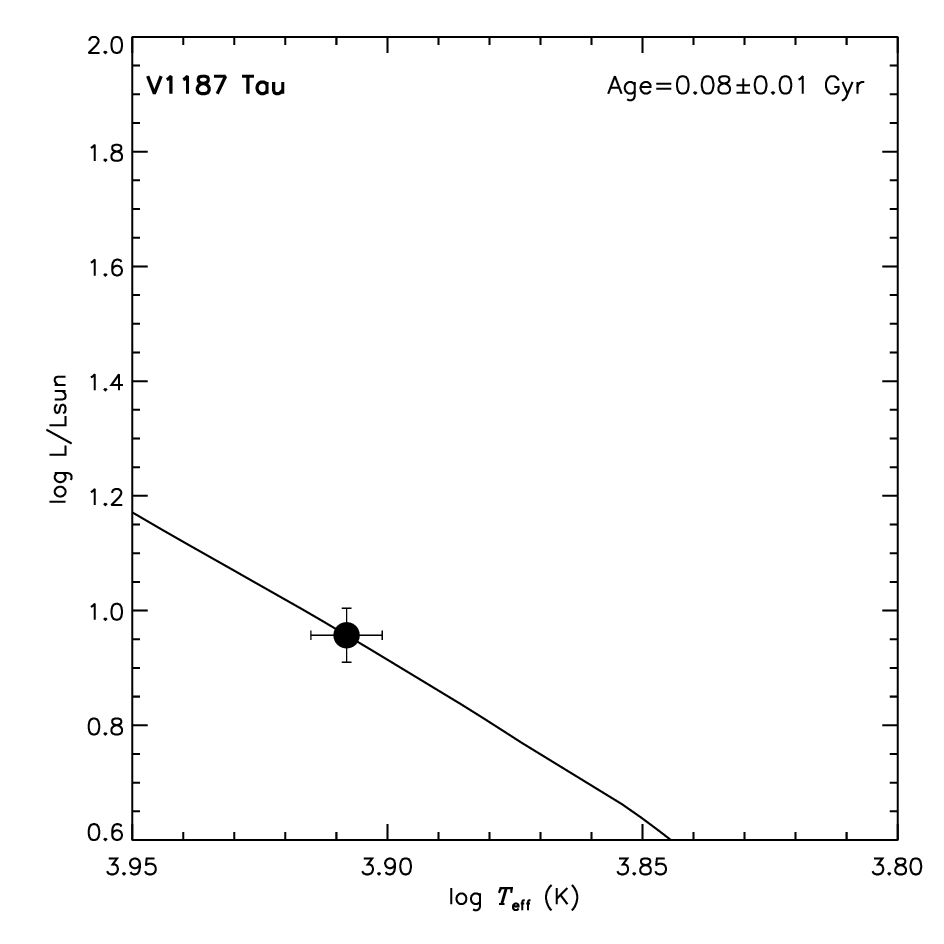}
  \end{minipage}
     \begin{minipage}[b]{0.4\textwidth}
  \includegraphics[alt={HR diagram for AU Scl},height=4.7cm, width=1\textwidth]{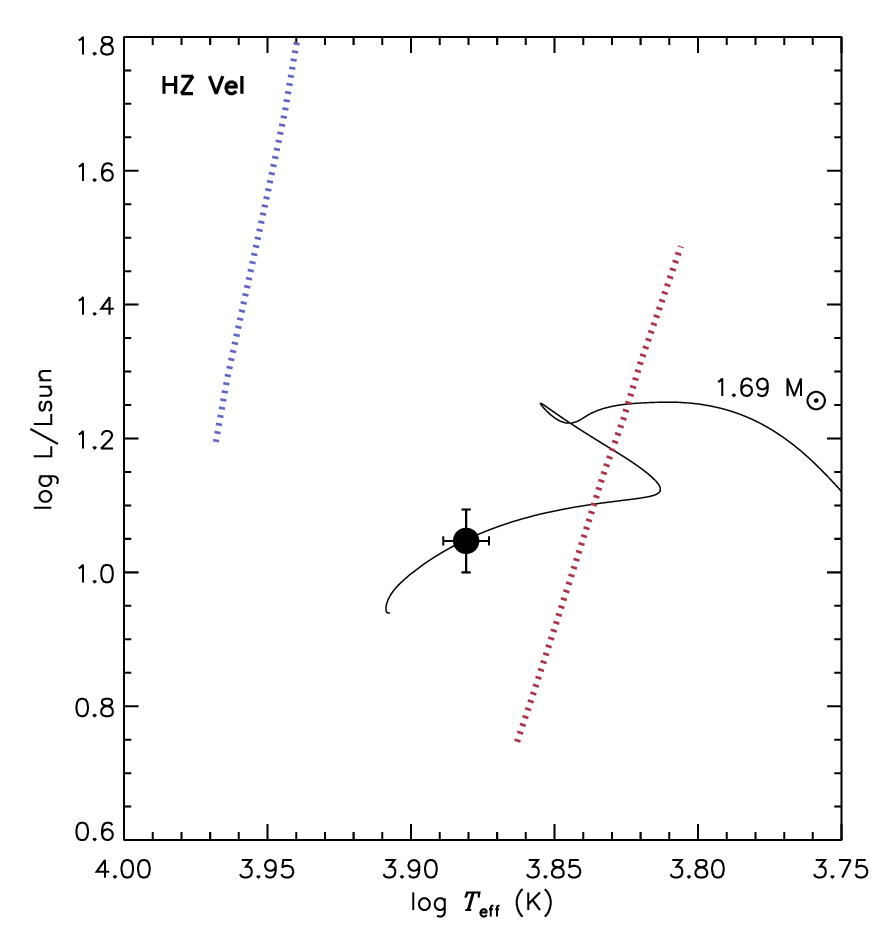}
  \end{minipage}
     \begin{minipage}[b]{0.4\textwidth}
  \includegraphics[alt={Isochrone for HZ Vel},height=4.7cm, width=1\textwidth]{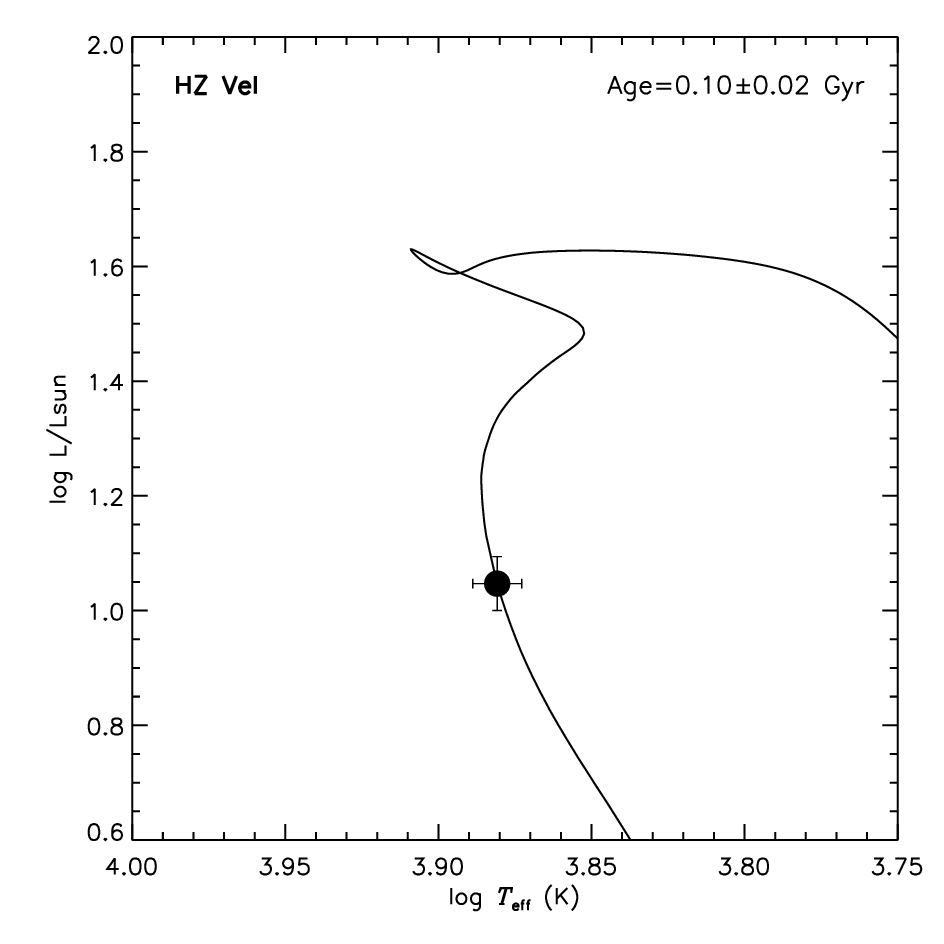}
  \end{minipage}
     \begin{minipage}[b]{0.4\textwidth}
  \includegraphics[alt={HR diagram for V527 Car},height=4.7cm, width=1\textwidth]{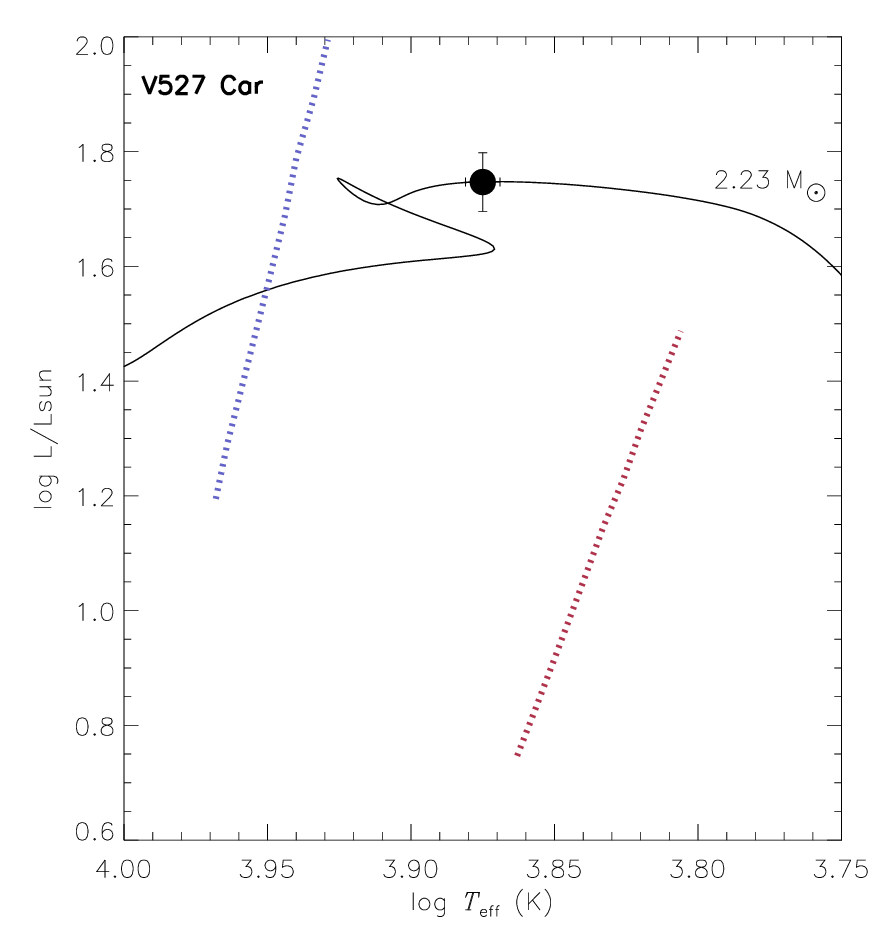}
  \end{minipage}
     \begin{minipage}[b]{0.4\textwidth}
  \includegraphics[alt={Isochrone for V527 Car},height=4.7cm, width=1\textwidth]{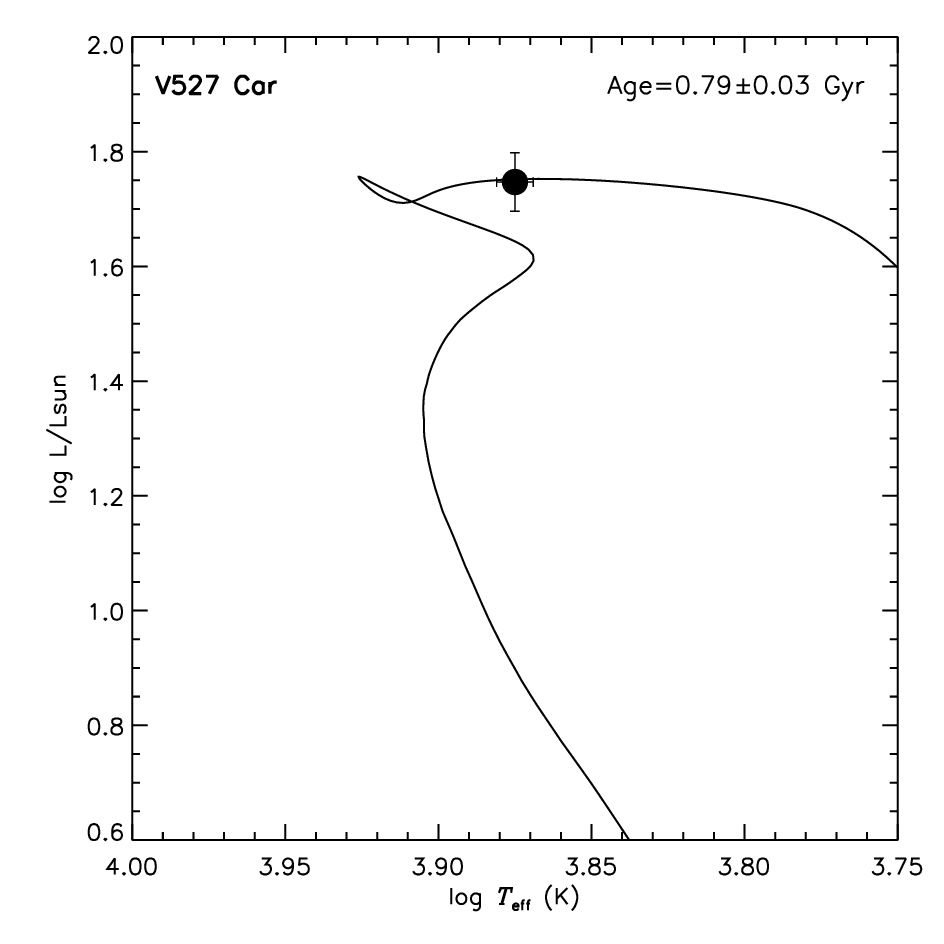}
  \end{minipage}
   \caption{Evolutionary tracks (left panels) and isochrones (right panels) for targets. The evolutionary tracks and isochrones were generated considering the Fe/H of each targets. The dotted lines represent the cool (red one) and hot (blue one) observational borders of \ds\, instability strip \citep{2019MNRAS.485.2380M}. }  \label{evo_isc}
\end{figure*}

\setcounter{figure}{4}
\begin{figure*}
 \centering
 \begin{minipage}[b]{0.4\textwidth}
  \includegraphics[alt={HR diagram for DP UMa},height=4.7cm, width=1\textwidth]{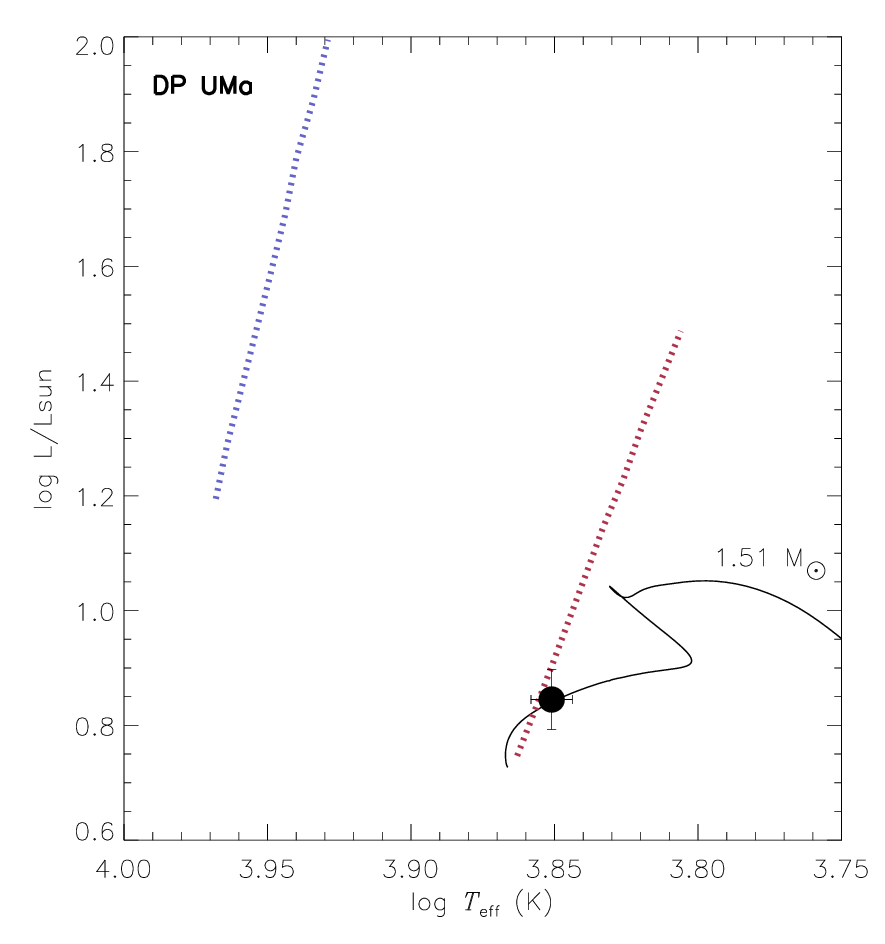}
 \end{minipage}
 \begin{minipage}[b]{0.4\textwidth}
  \includegraphics[alt={Isochrone for DP UMa},height=4.7cm, width=1\textwidth]{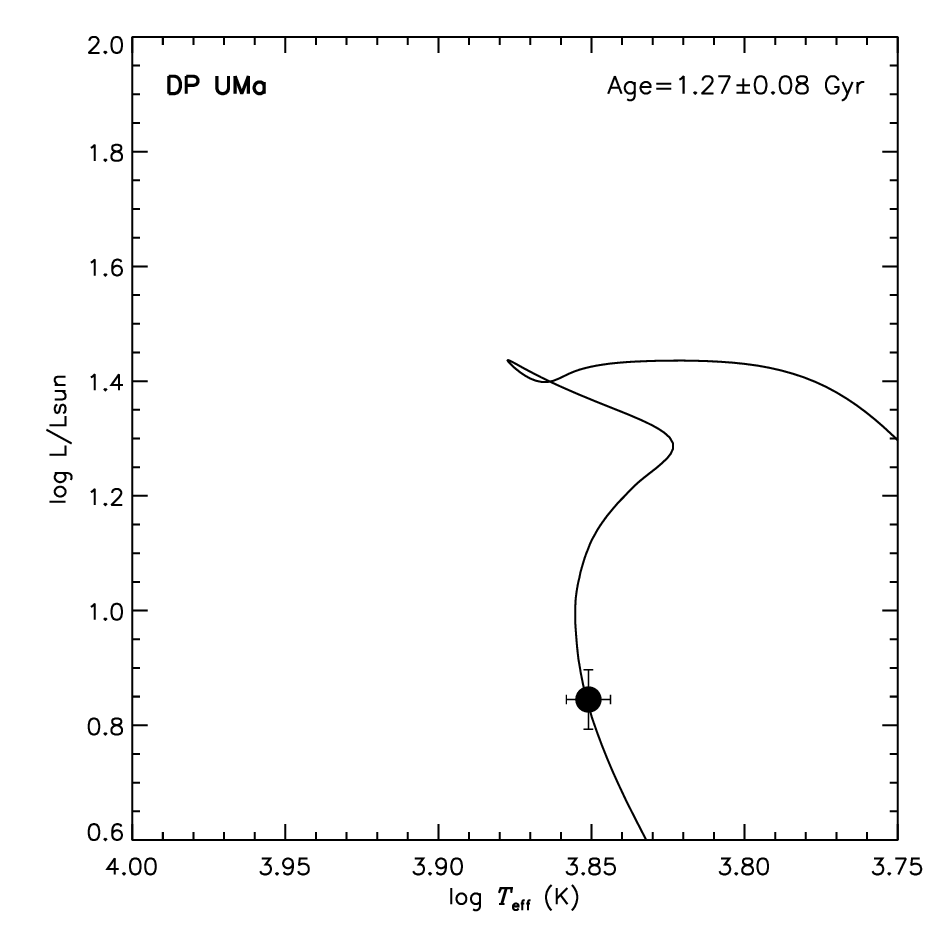}
  \end{minipage}
   \begin{minipage}[b]{0.4\textwidth}
  \includegraphics[alt={HR diagram for KU Com},height=4.7cm, width=1\textwidth]{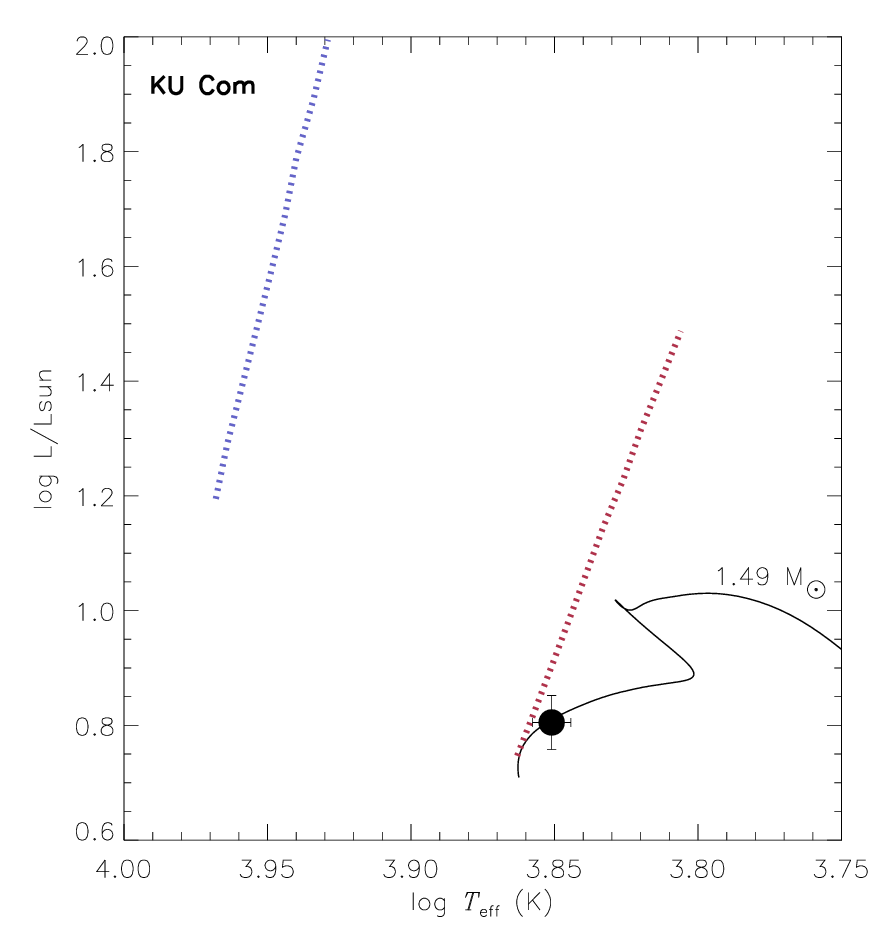}
  \end{minipage}
   \begin{minipage}[b]{0.4\textwidth}
  \includegraphics[alt={Isochrone for KU Com},height=4.7cm, width=1\textwidth]{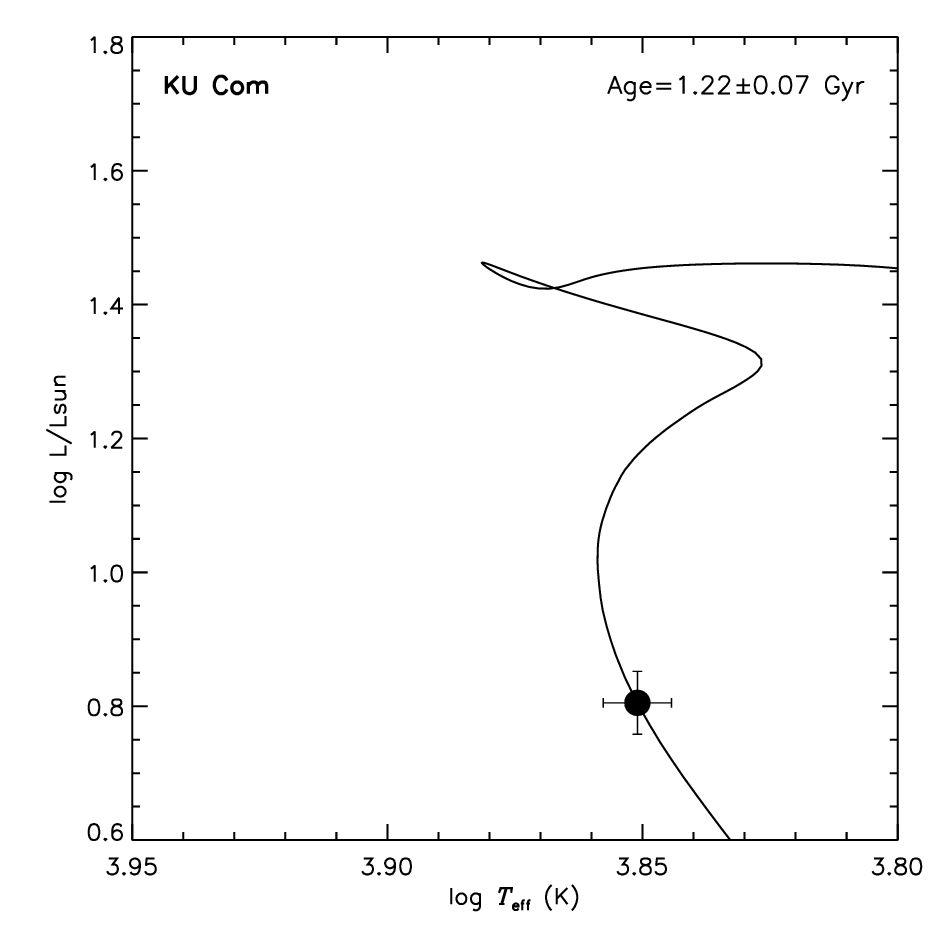}
  \end{minipage}
     \begin{minipage}[b]{0.4\textwidth}
  \includegraphics[alt={HR diagram for MX Vir},height=4.7cm, width=1\textwidth]{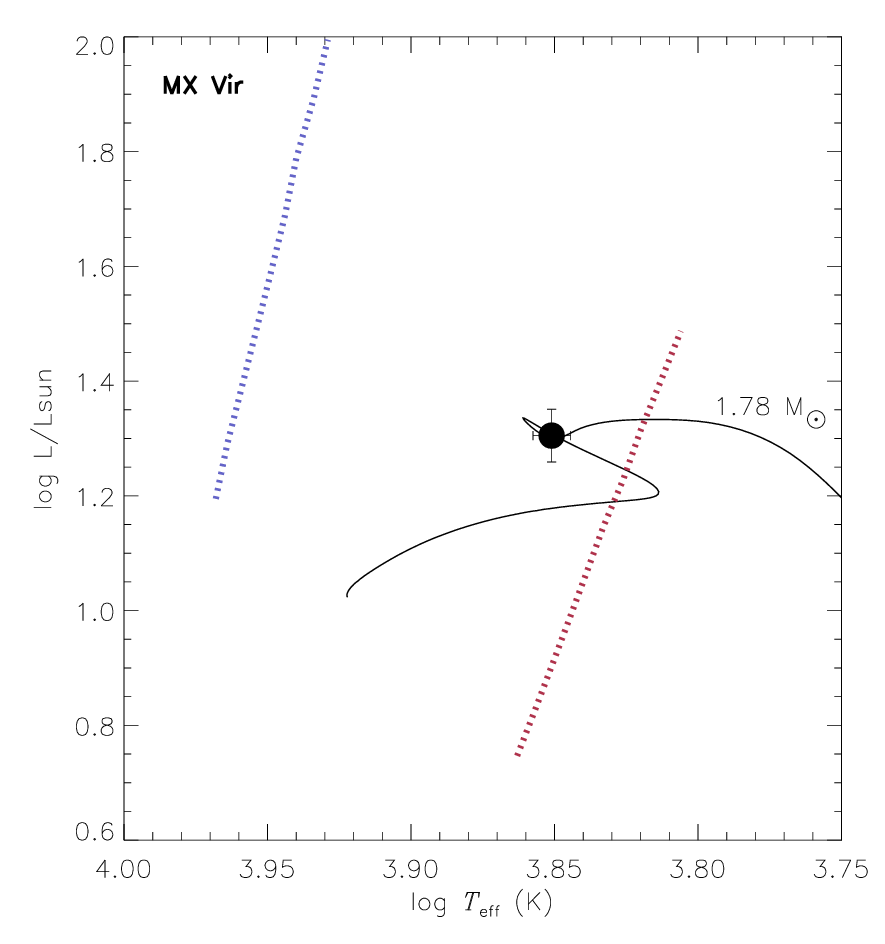}
  \end{minipage}
     \begin{minipage}[b]{0.4\textwidth}
  \includegraphics[alt={Isochrone for MX Vir},height=4.7cm, width=1\textwidth]{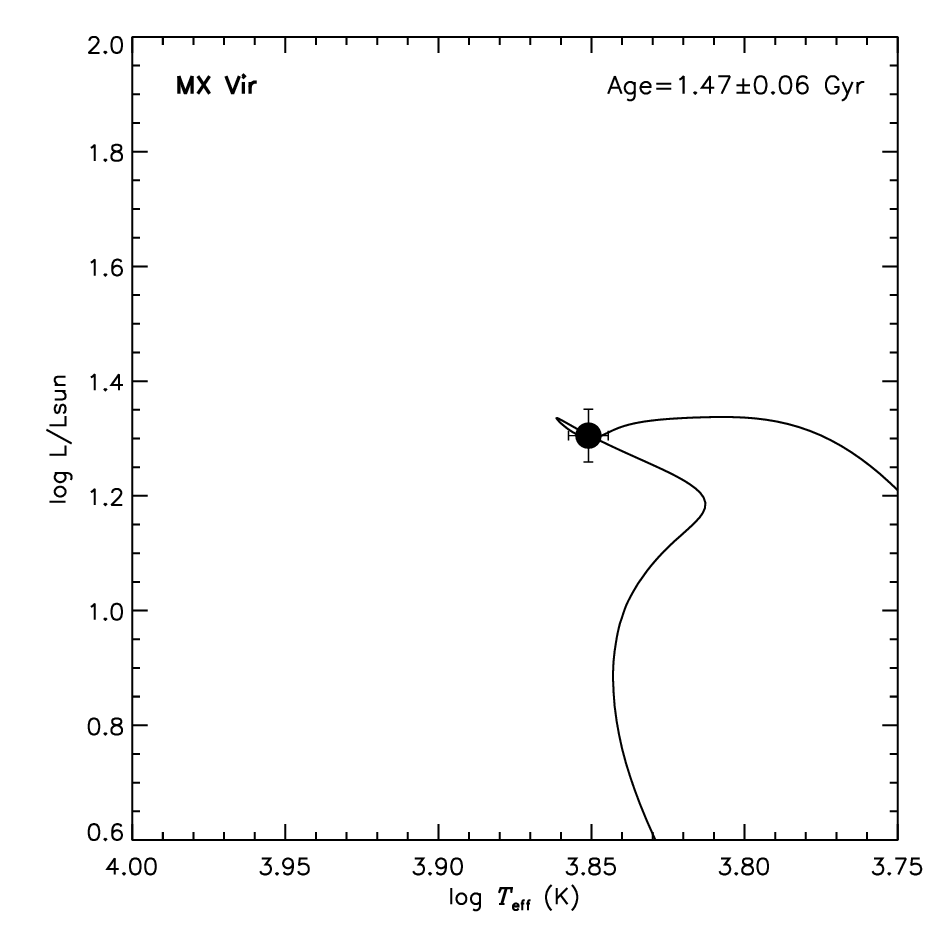}
  \end{minipage}
     \begin{minipage}[b]{0.4\textwidth}
  \includegraphics[alt={HR diagram for IO Lup},height=4.7cm, width=1\textwidth]{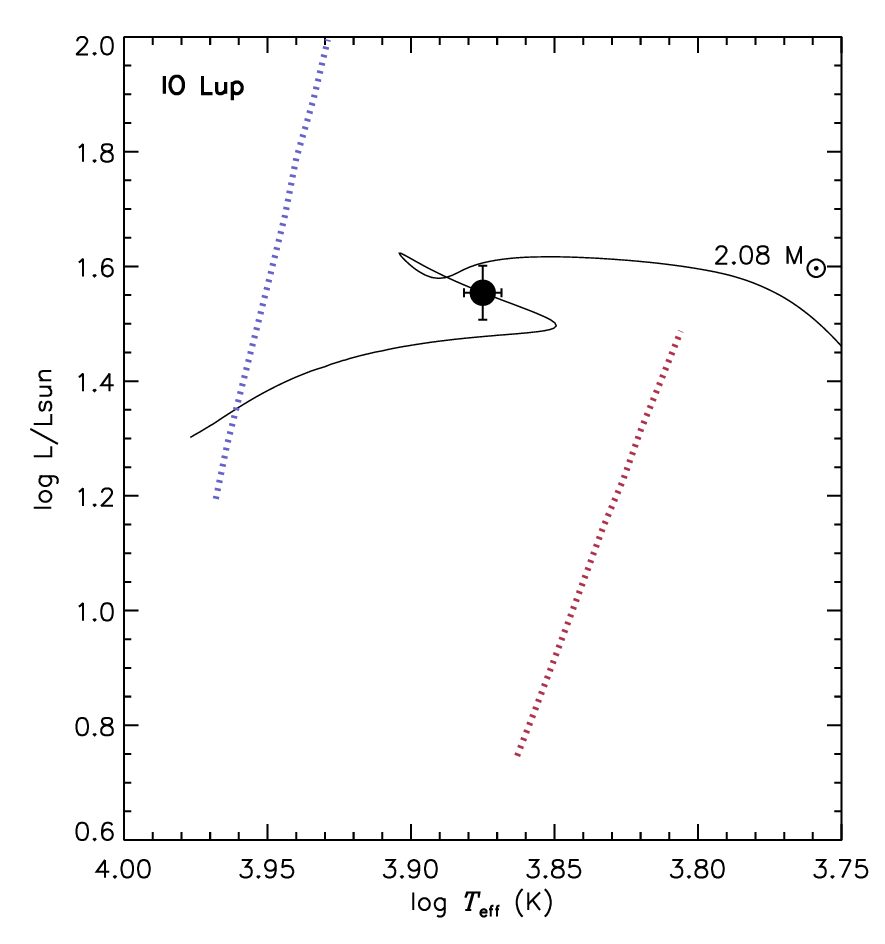}
  \end{minipage}
     \begin{minipage}[b]{0.4\textwidth}
  \includegraphics[alt={Isochrone for IO Lup},height=4.7cm, width=1\textwidth]{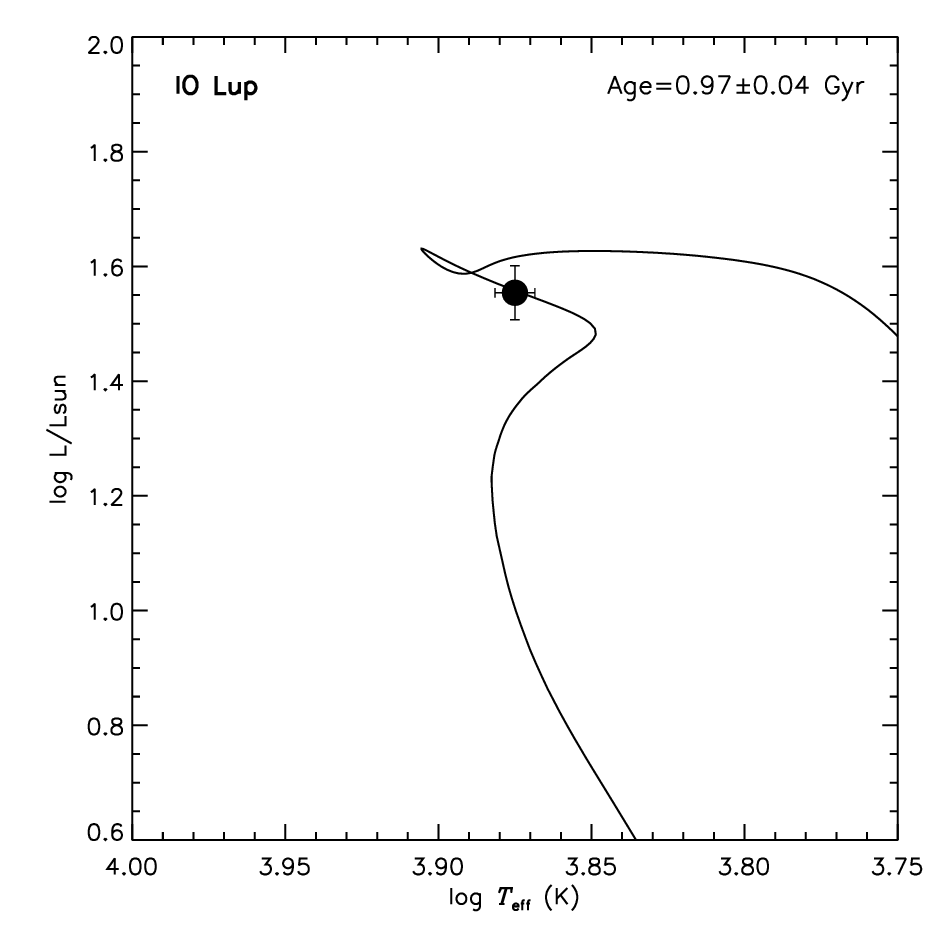}
  \end{minipage}
     \begin{minipage}[b]{0.4\textwidth}
  \includegraphics[alt={HR diagram for UV PsA},height=4.7cm, width=1\textwidth]{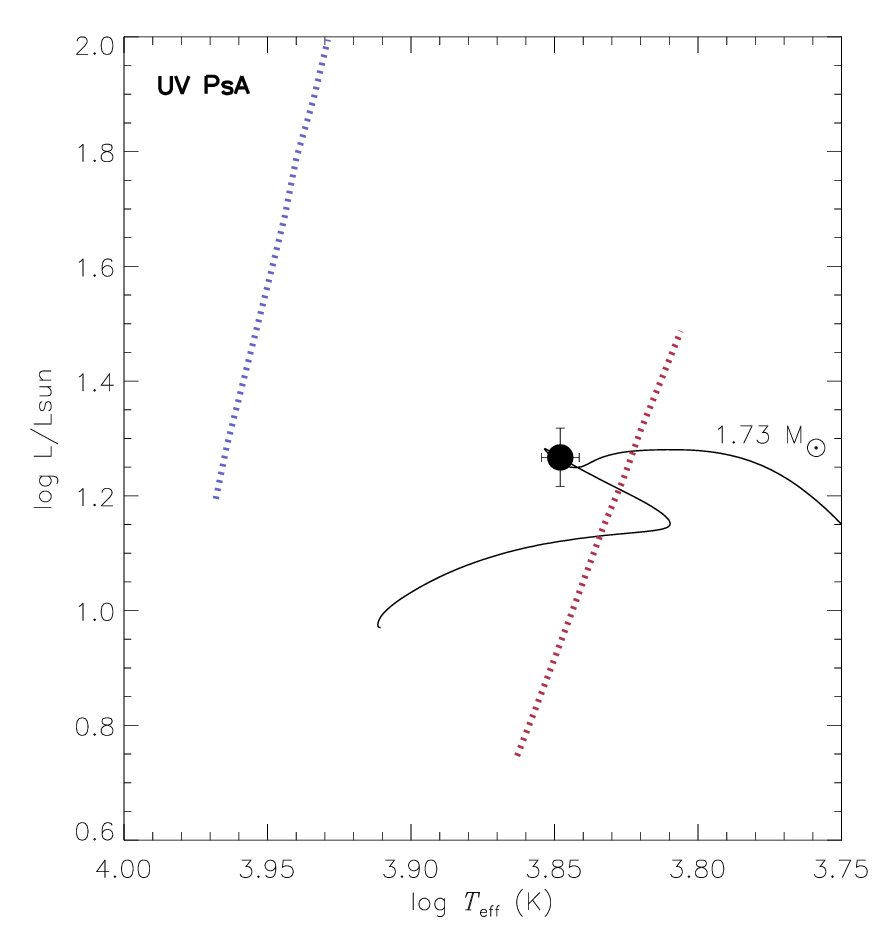}
  \end{minipage}
     \begin{minipage}[b]{0.4\textwidth}
  \includegraphics[alt={Isochrone for UV PsA},height=4.7cm, width=1\textwidth]{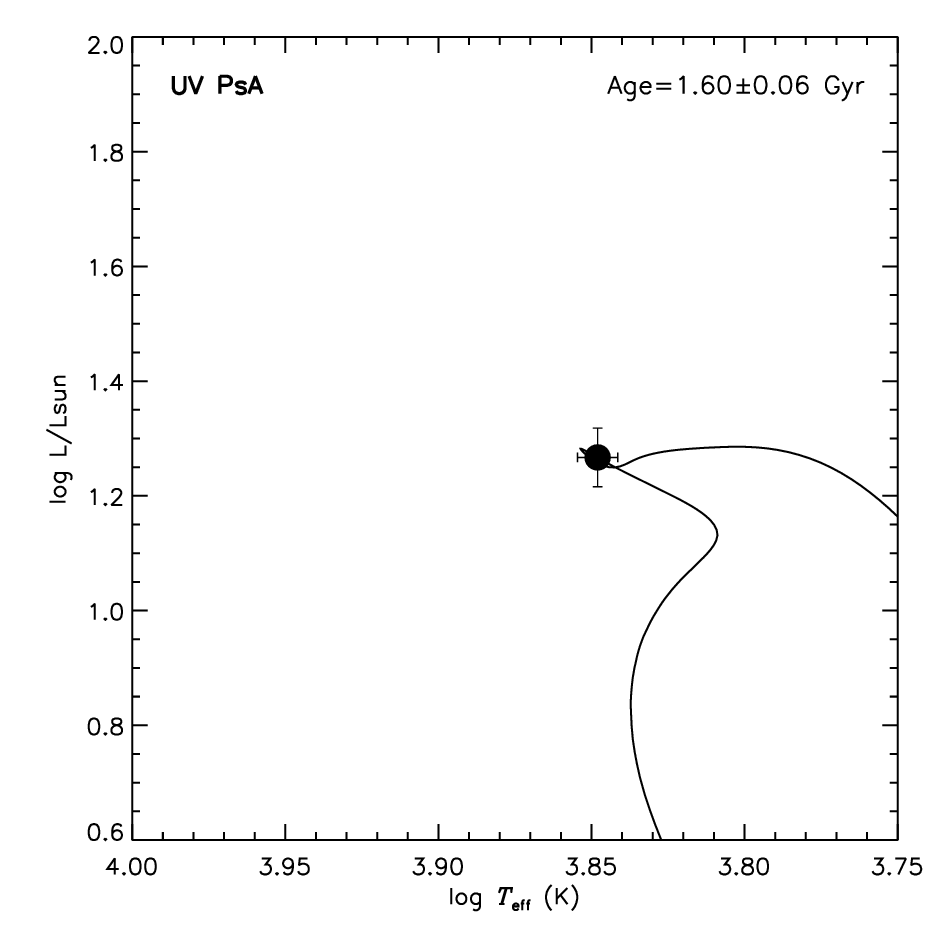}
  \end{minipage}
   \caption{Continuation.}\label{evo_isc2}
   \end{figure*}

\section{Discussion and Conclusion}

In this study, we examined ten targets identified in the literature as \ds\, variables and chemically peculiar stars (see Appendix). Although some of the selected stars have been previously studied, a detailed chemical abundance analysis has not been conducted for all of them. Therefore, we first revised their spectral classifications, determined their atmospheric parameters and \vsini\, values, and derived the abundances of individual elements. Using this information, we analyzed the chemical peculiarity properties of these targets.

We first carried out spectral classification for all targets and, as a result, found that only three stars, AU\,Scl, FG\,Eri, and HZ\,Vel are likely chemically peculiar. However, the other stars were found to be chemically normal based on our spectral classification, which contradicts their classifications in the literature (see Table\,\ref{table1}). Since the spectral classification was performed manually and is inherently subjective, we carried out a detailed chemical abundance analysis to confirm the chemical composition of the stars. In advance to chemical abundance analysis, the atmospheric parameters of the targets were determined. Most of our targets have atmospheric parameter estimates available in the literature. For example, five of them, AU\,Scl, FG\,Eri, MX\,Vir, IO\,Lup, and UV\,PsA, were analyzed by \citet{2016AJ....152..207R}, who investigated high-resolution HARPS spectra of a large sample of stars to determine atmospheric parameters (for details, see Appendix). We also used HARPS spectra in our spectral analysis. However, the \logg\ and [Fe/H] values obtained by \citet{2016AJ....152..207R} and those derived in this study are mostly inconsistent with each other. This discrepancy may result from differences in the methods and models used in both studies, as well as the fact that their analysis involved a significantly larger number of stellar spectra.  

For V1187\,Tau, the atmospheric parameters based on spectral analysis were reported by \citet{2018ApJ...857...93D} as \teff\,=\,8031\,K, \logg\ = 4.0, and [Fe/H] = -0.17. When compared to our results, they are in agreement within the error bars, except for the [Fe/H] value. For DP\,UMa, \citet{2014MNRAS.441.1669C} provided the spectral parameters as \teff\,=\,7200,K, \logg\,=\,3.6, and \vsini\,=\,72\,\kms. These values are generally consistent with ours, although the \logg\, value shows a significant difference (our value is 4.2). For the final target, KU\,Com, \citet{2011A&A...531A.165P} reported \teff\,= 7360\,K, \logg\,=\,3.99, and [Fe/H]\,=\,-0.05, all of which agree with our findings within uncertainties while [Fe/H] differs from ours ([Fe/H]=-0.21). The differences between our and literature results could be the reason of followed methods, used atmospheric models, and line list. For the other objects, V527\,Car and HZ\,Vel, no spectral atmospheric parameter estimates are available in the literature.

After the estimation of atmospheric parameters the chemical abundances of the targets were derived. As given in the Appendix, most of the targets have been classified as Am stars in the literature, which exhibit an overabundance of iron-peak elements \citep{2009ssc..book.....G} and some of the targets have also been identified as metal-weak $\lambda$\,Bootis stars, characterized by under abundances of iron-peak elements \citep{2009ssc..book.....G}. Therefore, considering all chemically peculiar star types along with our chemical abundance results obtained in this study, we reclassified the peculiarities of the stars. Based on our analysis, we identified three chemically peculiar stars among our targets. AU\,Scl and FG\,Eri were classified as Am stars, while HZ\,Vel exhibited typical $\lambda$\,Bootis-type chemical properties. The remaining objects were classified as normal stars based on their abundance distributions. Among the target stars, only two objects AU\,Scl and FG\,Eri were identified as metallic A stars based on our spectroscopic analysis. AU\,Scl is located within the region of the H-R diagram suggested for pulsating Am stars by \citet{2017MNRAS.465.2662S} and \citet{2024A&A...690A.104D}, whereas FG\,Eri deviates from this common location expected for metallic A stars. Notably, FG\,Eri exhibits a relatively high pulsation constant. When considered together with the distinct position of FG\,Eri on the H-R diagram and its chemical peculiarities, these findings make it an interesting case for further theoretical investigation. FG\,Eri may represent a transition object or an outlier that could provide constraints on models of pulsation in metallic A stars.

The pulsation structures of the targets were also examined, and their pulsation types were identified. According to the oscillation analysis, most of the stars were classified as \ds\, variables, consistent with the literature classification (see Appendix) and some were identified as \ds\,-\gd\, hybrids. The pulsation frequencies of eight of our targets (expect for V1187\,Tau and IO\,Lup) were examined by \citet{2022MNRAS.516.2080B} using TESS data. They provided the dominate \ds\, frequencies for each targets and these results are consistent with our findings. For V1187\,Tau and IO\,lup the dominate frequencies were given as 53.18\,d$^{-1}$ \citep{2023ApJ...946L..10B} and 13.40\,d$^{-1}$ \citep{2019ApJS..244...15M}, respectively in their latest studies. While the result for V1187\,Tau is matching with ours, for IO\,Lup it differs because this study is basing on the grand-based photometric observations.

In addition to estimating the pulsation frequencies for each target, we also determined the pulsation modes of the dominant (highest amplitude) frequency using the Q values and the log\,P${puls}$\,-\,M$_{V}$ relationship. Furthermore, we presented the positions of the stars on the Hertzsprung–Russell diagram and within the \ds\, instability strip. All targets were found to lie within or near the cool edge of the \ds\, instability strip. While some of the hybrid stars (HZ\,Vel, DP\,UMa, KU\,Com) are located close to the cool edge as their suggested region, others are situated in the central region of the instability strip.

We estimated fundamental stellar parameters such as $M$ and age for our targets. To achieve this, we first calculated the $L$ for each target and then generated evolutionary tracks and isochrones based on the corresponding [Fe/H] values. By identifying the best-fitting evolutionary tracks and isochrones that match the current positions of the stars on the Hertzsprung–Russell diagram, we determined the $M$ and age values. 



We conducted a detailed analysis of ten possible chemically peculiar stars and found that most of them are, in fact, chemically normal. Such studies highlight the importance of spectroscopic analysis in reliably classifying stars as chemically peculiar. Even a basic spectral classification can provide valuable insights into chemical peculiarities; however, a more robust identification of chemically peculiar stars can be achieved through comprehensive spectroscopic investigations. Moreover, analyzing such stars is crucial for understanding the pulsation structures of \ds\, stars and their relationship with the chemical abundances of individual elements, especially through spectral diagnostics.

Our findings further support recent results indicating that chemically peculiar A stars can indeed pulsate in the \ds\, domain, challenging earlier assumptions that atomic diffusion and helium settling would suppress such pulsations. The identification of AU\,Scl and FG\,Eri as metallic-line A stars with \ds\, pulsations is consistent with the conclusions of \citet{2024A&A...690A.104D}, who found that many pulsating metallic A and F stars cluster near the red edge of the instability strip on the H–R diagram. Our spectroscopic results confirm AU\,Scl's position in this expected region, while FG\,Eri appears to deviate, which, along with its relatively large pulsation constant, renders it an interesting outlier that may provide constraints for future theoretical models. Understanding chemically peculiar stars with confirmed pulsational properties may contribute significantly to stellar structure and evolution theories, particularly in the transition region between chemically normal and peculiar A-type stars. Therefore, our work not only reassesses the classifications of several candidates but also adds spectroscopically verified examples to the growing population of pulsating CP stars, strengthening the empirical basis for such theoretical efforts.

\section*{Acknowledgements}

This study has been supported by Canakkale Onsekiz Mart University Research foundation through project FBA-2020-3219. We thank the referee for the review and remarks. The TESS data presented in this paper were obtained from the Mikulski Archive for Space Telescopes (MAST). Funding for the TESS mission is provided by the NASA Explorer Program. This work has made use of data from the European Space Agency (ESA) mission Gaia (http://www.cosmos.esa.int/gaia), processed by the Gaia Data Processing and Analysis Consortium (DPAC, http://www.cosmos.esa.int/web/gaia/dpac/consortium). Funding for the DPAC has been provided by national institutions, in particular, the institutions participating in the Gaia Multilateral Agreement. This research has made use of the SIMBAD database, operated at CDS, Strasbourg, France.

\newpage
\appendix \label{appendix}

\section{Literature information of targets}

\subsection{AU\,Scl}
AU\,Scl was first reported in the list of early-type stars near the South Galactic pole given by Slettebak and Brundage (1971), and the spectral type of AU\,Scl was given to be A7p in this list. \citet{1973AJ.....78..295G} discussed the spectroscopically peculiar stars near the South Galactic pole, and they noted that AU\,Scl could be a Peculiar A star of $\beta$ CrB type or late-type Am star. \citet{1989MNRAS.238.1077K} discovered \ds\, variability in AU\,Scl (= HD\,1097) for the first time and concluded that AU\,Scl is a classical Am star based on the spectral classification of A3/5mF0-F5 given by \citet{1982mcts.book.....H}. \citet{2011MNRAS.414..792B} listed the pulsating Am stars not in the Kepler field and gave the \teff\, and log(L/L$_{\odot}$) of AU\,Scl to be 7586\,K and 0.11, respectively based on the photometric colors. \citet{2011A&A...535A...3S} studied 1600 Am stars with SuperWASP data, and they gave the frequency, the amplitude, and the class of the pulsating Am star AU\,Scl as 15.5491\,d$^{-1}$, 2.01\,mmag and \ds, respectively. \citet{2016AJ....152..207R} formed a large database of the physical parameters and variability indicators of the variable and active stars using HARPS spectra, including AU\,Scl. In this study they estimated the atmospheric parameters of the star to be \teff\,=\,6595\,K, \logg\,=\,3.70, and Fe/H\,=\,0.31. \citet{2022MNRAS.516.2080B} studied \ds\, stars and their period-luminosity relation using TESS and Gaia DR3 data. They gave the dominate frequency of AU\,Scl as 16.55\,d$^{-1}$. 

\subsection{FG\,Eri}
The first study of FG\,Eri was given by \citet{1980ApJS...44....1T}. \citet{1999A&AS..135..503G} determined radial velocities of B8-F2 type stars observed by the Hipparcos satellite. They gave the \teff\, and spectral type of FG\,Eri as 8500\,K and A5V, respectively. \citet{2001A&A...373..625P} presented a spectroscopic survey for $\lambda$ Boo type stars listing the galactic stars, of which one of the list stars was FG\,Eri with spectral type kF0hA5mF0V. \citet{2002A&A...393..897R} determined the rotational velocities of A-type stars using the spectra of 525 B8 to F2-type stars collected at ESO, and they derived the \vsini\, of FG\,Eri as 95\,$\pm$\,10\,\kms. \citet{2002ChJAA...2..441X} plotted the \ds\, stars on the Hertzsprung–Russell diagram to examine the theoretical red edge. They found some stars which have higher luminosity and located beyond the \ds\, instability strip, one of which is FG\,Eri. They excluded these stars from their instability strip investigation. \citet{2012A&A...546A..61D} determined the radial velocity errors for the Hipparcos-Gaia HTPM Project and gave the \teff\, and \vsini\, of FG\,Eri as 7873\,K and 107.2 \kms, respectively. \citet{2016AJ....152..207R} formed a large database of the physical parameters and variability indicators of the variable and active stars using HARPS spectra, including FG\,Eri in that study \teff, \logg, and Fe/H were determined to be 8986\,K, 3.93 and 0.71, respectively. \citet{2019MNRAS.490.4040A} determined stellar parameters and variability type of the 2-min cadence TESS \ds\, and \gd\, pulsators. They defined FG\,Eri’s variability type as developed \ds\,. \citet{2020A&A...638A..59B} analyzed the \ds\, stars using TESS data to propose the frequency at the maximum power, and classified the star FG\,Eri to be \ds\,/\gd\, hybrid. The dominant frequency measured with TESS of FG\,Eri was given in \citet{2022MNRAS.516.2080B} to be 6.29\,d$^{-1}$. 

\subsection{V1187\,Tau}

V1187\,Tau belongs to the open cluster Cl Melotte 22 (= Pleiades). \citet{1965ApJ...142.1604A} gave the spectral type of the star as A5\,V and noted that the radial velocity of V1187\,Tau may vary slowly with time. \citet{1968ApJ...152..483C} analyzed the chemical abundances of V1187\,Tau, and eight stars in the Pleiades, and concluded that measured Sc/Sr ratios of these stars support the hypothesis that stars showing Sc/Sr ratios near unity are those with normal metal abundances. \citet{1975PDAO...14..319P} observed a total of 115 stars in the Pleiades at DAO spectroscopically to investigate the binary friction, cluster membership and they obtained radial velocity value for V1187\,Tau. \citet{1978PASP...90..201A} presented the spectral classifications and absolute magnitudes of stars in the Pleiades and noted that V1187\,Tau was a marginal Am star. \citet{1991AJ....101.1495M} measured the radial velocities of late O-B type and early A type stars, including V1187\,Tau. 
\citet{1999MNRAS.309.1051K} detected a pulsation with a period of about 30 min. for V1187\,Tau, and concluded that V1187\,Tau may be a marginal Am star in a binary system. \citet{2002A&A...382..556F} presented the results obtained in the STEPHI X multi-site campaign for two \ds\, stars (V624\,Tauri and HD\,23194\,=\,V1187\,Tau). They found to be a multi-periodic, non-radial pulsator with two modes of oscillations for V1187\,Tau. \citet{2006MmSAI..77..455F} performed a seismological analysis of six \ds\, stars, one of which was V1187\,Tau, belonging to the Pleiades cluster to identify their frequency modes. They found two pulsation frequencies for V1187\,Tau as 46.10 and 49.65\,d$^{-1}$. \citet{2011A&A...531A.165P} re-determined the atmospheric parameters of the stars in the medium spectral resolution library MILES containing V1187\,Tau. They found the atmospheric parameters to be \teff\,=\,8031\,K, \logg\,=\,4.00 and Fe/H\,=\,-0.17. \citet{2023ApJ...946L..10B} studied a total of 89 A-F type members of open cluster Pleiades, and found the highest amplitude frequency as 53.18\,d$^{-1}$ using TESS data. 

\subsection{HZ\,Vel}

In the literature, the first information about HZ Vel was given in \citet{1972MNRAS.160..155S} study based on ubvy and H$_{\beta}$ photometry. \citet{1977MmRAS..84..101B} searched the southern $\beta$ CMa stars and noted that HZ\,Vel was a new \ds\, star with pulsation period of 0.087 days. \citet{1977A&A....61..563H} applied cluster analysis method and redetermined the spectral classification of HZ\,Vel to be A7-F2\,III-V. \citet{1980A&A....85...93M} determined the spectral classification of the star HZ\,Vel to A8\,III using their algorithm based on uvby$\beta$ photometry and MK classification. \citet{1986A&A...155..371H} included HZ\,Vel to the list of suspected $\lambda$ Bootis stars in their search for these types of candidates. Hauck (1986) studied the metallics of 132 stars considered as A and F giant stars and expressed that HZ\,Vel could be assumed to be $\lambda$\,Bootis star regarding its low metallicity value. \citet{1994A&AS..106...21R} presented the updated list of \ds\, stars, including HZ\,Vel. \citet{1997A&AS..123...93P} determined the distance and absolute magnitude of HZ\,Vel, and remaining $\lambda$ Bootis type program stars to derive the evolutionary state of these stars using Hipparcos data. \citet{1998A&A...335..533P} gave the list of pulsating $\lambda$ Bootis stars, including HZ\,Vel. \citet{1999A&A...345..597P} derived stellar parameters and abundances of their sample stars with HZ\,Vel. HZ\,Vel was included in the study of \citet{1999A&A...349..521F} as a $\lambda$ Boo candidate star. \citet{2002MNRAS.336.1030P} reobserved HZ\,Vel and found two frequencies 14.80 and 15.99\,d$^{-1}$ using their observations. \citet{2002A&A...393..897R} measured the \vsini\, of the northern star HZ Vel to be 45\,\kms. \citet{2004A&A...425..615F} studied the sample of $\lambda$\,Boo stars, which shows composite spectra, given by \citet{2003A&A...412..447G}, and classified HZ\,Vel as the suspected double. \citet{2017MNRAS.466..546M} gave the parameters of 172 $\lambda$\,Boo stars and candidates using Gaia DR1/Hipparcos parallaxes, and derived \teff\,, log(L/L$_{\odot}$) and variability type of HZ\,Vel to be 7270\,$\pm$\,85 K, 1.06 and \ds\,, respectively. 
\citet{2020A&A...638A..59B} derived the power spectra of \ds\, stars with TESS and determined the \teff\, (7400\,K) and variability type ($\delta$ Scuti/$\gamma$ Doradus hybrid) of HZ Vel. \citet{2020MNRAS.495.1888M} derived the pulsation features of southern $\lambda$\,Boo stars using TESS data, and gave the atmospheric parameters of HZ\,Vel as \teff\,=\,7270\,$\pm$\,145 K, \logg\,=\,4.03\,$\pm$\,0.05, [Fe/H]\,=\,0.00-0.17\,$\pm$\,0.15 basing on the photometric investigations. The dominate frequency of the stars was given as 14.34\,d$^{-1}$ in the study of \citet{2022MNRAS.516.2080B}. 

\subsection{V527\,Car}

V527\,Car was recorded for the first time by \citet{1975mcts.book.....H}, in which the spectral type of the star was given as A3mA7-A9. \citet{1997IBVS.4513....1D} investigated true and possible contact binaries in the Hipparcos catalog and commented the star V527\,Car as pulsating or EW; visual binary. \citet{1998A&AS..133....1P} gave new variable chemically peculiar stars using the extensive Hipparcos Variability Annex, one of which is V527 Car. \citet{1999MNRAS.309.1051K} found that V527\,Car is an evolved Am (Puppis) star with a 5.1-h periodicity, and identified the mode of its pulsation as non-radial with l = 2 based on the multicolor photometry. The dominant frequency measured of V527\,Car was given in the study of \citet{2022MNRAS.516.2080B} to be 4.68\,d$^{-1}$.

\subsection{DP\,UMa}

The target star DP\,UMa is one of the peculiar stars with spectral type A5 \citep{1932ApJ....75...46M, 1959AAHam...5..105S}. Eggen (1976) discussed the photometry and kinematics of Am stars and ultrashort-period cepheids (USPC) in the HR catalog, giving DP UMa’s spectral type to be A7m. \citet{1978ApJ...221..869K} first announced the discovery of pulsation of DP\,UMa with other Am star HR 8210. \citet{1979MNRAS.186..575K} expressed that the frequency of DP\,UMa changes on a time-scale as short as 1 day. \citet{1982SvA....26..666T} estimated the radial pulsation modes for \ds\, stars from \citet{1979PASP...91....5B} list and gave the oscillation mode of DP\,UMa. \citet{1994MNRAS.267.1045W} reported the range of temperature classes covered by the spectral type of DP\,UMa is rather small (kA4hA6mA7), implying that the star DP\,UMa is quite a marginal Am star. \citet{1997BaltA...6..499B} made classification of Population II stars in Vilnius photometric system, and identified the class of DP\,UMa as the optical binary. \citep{2008AJ....136.1061B} measured the rotational velocities of 118 \ds\, variables using DAO spectra and gave the \vsini\,=\,70 km/s for DP\,UMa. \citet{2014MNRAS.441.1669C} presented a detailed analysis of eight Am stars, and obtained the \teff, \logg\, and \vsini\, of DP\,UMa as 7200\,K, 3.6, and 72\,\kms. They also reported the binary feature, belonging to the Am class and the presence of pulsation for DP\,UMa. \citet{2018ApJ...857...93D} derived the stellar parameters (\teff\,=\,7514 K, \logg\,=\,3.5, \vsini\,=\,70 km/s for DP\,UMa) of their program stars using the spectra obtained at McDonald observatory. \citet{2022MNRAS.516.2080B} obtained some properties of 372 \ds\, stars, and gave the dominate frequency value to be 18.07\,d$^{-1}$ for DP\,UMa.

\subsection{KU\,Com}

KU\,Com is a member of open cluster Cl Melotte 111 \citep{1950ApJ...111..414E, 1952ZA.....31..236B, 1955ApJ...122..209J}, and is a \ds\, type variable star with A7m spectral type \citep{1959AAHam...5..105S}. \citet{1991AJ....101.2177S} reported four new probable \ds\, stars, one of which was KU\,Com, and concluded that KU\,Com could be a metallic line star, thus might be a member of the $\delta$ Del group. \citet{1994A&AS..106...21R} presented the extensive list of \ds\, stars covering the basic properties of KU\,Com. 
\citet{1997A&A...323..901H} gave metal abundances of KU\,Com and remaining program stars in young galactic clusters. 
\citet{1999ApJ...521..682A} studied the evolution of binary systems in five open clusters of various ages, and derived the constant velocity for KU\,Com which is one of the observed stars in the Coma cluster.  \citet{2000A&A...354..216B} gave the third in a paper series about the abundances of the elements Li, Al, Si, S, Fe, Ni and Eu for A-stars, including KU\,Com, in open clusters of different ages. \citet{2003AJ....126.2408M} corrected the Hipparcos parallaxes of Coma Berenices and NGC 6231 open clusters, giving the recomputed parallax of 12.35(0.94) for KU\,Com. \citet{2004IAUS..224..209M} gave the abundances obtained by synthesizing AURELIE spectra for KU\,Com and other A and F dwarfs in Coma Berenices. \citet{2007MNRAS.374..664C} presented the set of stellar atmospheric parameters of 985 stars counting KU\,Com found in a new spectral stellar library MILES. \citet{2008A&A...479..189G} derived the chemical composition (abundances relative to hydrogen and to the solar value) of KU\,Com with spectral type Am/kA7hF0mF0. \citet{2011MNRAS.414.1227M} obtained [Mg/Fe] measurement for KU\,Com. Their work is based on understanding stellar atmospheres and stellar populations in galaxies and star clusters using the stars in the MILES spectral library. \citet{2011A&A...531A.165P} determined the MILES atmospheric parameters \teff, \logg, Fe/H of KU\,Com as 7360\,K, 3.99, and -0.05, respectively. \citet{2014A&A...566A.132S} aimed to obtain accurate ages for the seven nearby open clusters, each of which contains at least one magnetic Ap, one of these stars being KU\,Com, or Bp star. \citet{2017AJ....153..257O} cataloged the candidate commoving pairs of stars in the Tycho-Gaia Astrometric Solution (TGAS) which is the primary sample of the Gaia Data Release 1. Barac et al. (2022) provided a catalog of bright 372 \ds\, stars and their period-luminosity relation using TESS and Gaia DR3 data, and determined the dominant frequency as 19.91\,d$^{-1}$. \citet{2023A&A...675A..28O} applied their membership algorithm to the nearly 45 million Gaia DR3 sources and found 302 candidate members, one of which is KU\,Com, in the cluster Coma Berenices.

\subsection{MX\,Vir}

The first study of MX\,Vir was aimed at determining its spectral type \citep{1979PASP...91..176A}. They re-defined the spectral type of stars with unusual photometric indices and found that MX\,Vir is a star with F5\,Vp (Sr, Cr, Eu strong; Ca weak) spectral type. \citep{1996PASP..108..772P} computed color excesses and distances for southern A-early F stars, including MX\,Vir. \citet{1998A&AS..133....1P} identified the new variable chemically peculiar stars using the Hipparcos catalog, in which the spectral type of MX\,Vir was given as F3. \citet{2006A&A...449..281D} gave the spectroscopic study results of 37 candidate \gd\, stars, and they determined MX\,Vir as a \ds\, star with a F2\,II spectral type. \citet{2008A&A...478..487B} studied the southern \gd\, stars, spectroscopically, and gave the abundance amounts of MX\,Vir. \citet{2011A&A...530A.138C} presented improved astrophysical parameters for the Geneva-Copenhagen Survey (GCS). They gave the \teff\,, Fe/H and \logg\, of MX\,Vir to be 6803\,$\pm$\,136 K, 0.31 and 3.57, respectively. \citet{2016AJ....152..207R} presented spectroscopic information on the observed targets using HARPS spectra. In this study, MX\,Vir’s parameters, and variability type were given as \teff\,=\,7002 K, \logg\,=\,3.25, [Fe/H]\,=\,-0.06 and \ds\,. Mean \teff\, and \logg\, of MX\,Vir was given in \citet{2020A&A...638A..59B} study to be 7400\,K and 3.58, respectively. \citet{2022MNRAS.516.2080B} determined the dominat frequency to be 6.49\,d$^{-1}$ using TESS data. \citet{2024A&A...681A.107R} presented a new spectral library consisting of 3256 spectra containing 2043 stars, including MX\,Vir. 

\subsection{IO\,Lup}

IO\,Lup is one of the relatively least studied stars among our target stars. It was recorded in \citet{1996IBVS.4297....1P} which they chose as a comparison star for a survey to detect variability in $\lambda$ Bootis stars, because of IO\,Lup’s classification as Am star \citep{1982mcts.book.....H}. \citet{2002A&A...393..897R} measured \vsini\, of A-type stars in the northern and southern hemispheres. They gave the rotational velocity and spectral type of the star as 15\,km/s and A5mA5-F2. \citet{2016AJ....152..207R} determined the atmospheric parameters of IO\,Lup as \teff\,=\,7550 K, \logg\,=\,3.37, [Fe/H]\,=\,0.28 using the spectra obtained with the ESO echelle spectrograph HARPS. \citet{2019ApJS..244...15M} searched for periodic variations in the data obtained by Ring instruments located in South Africa and Australia, including IO\,Lup in their list and they gave a pulsation frequency value of 13.40\,d$^{-1}$ for IO\,Lup.

\subsection{UV\,PsA}

\citet{1982mcts.book.....H} gave the spectral type for UV PsA as F2\,III(m). 
\citet{1995MNRAS.277..217K} presented differential photometry in the VB wavebands for 11 \ds\, stars together with UV PsA, nine of which were not previously known to be variable stars. \citet{2011A&A...535A...3S} studied over 1600 Am stars using SuperWASP photometric data with a photometric precision of 1\,mmag. They gave the stellar parameters (Sp. Type\,=\,F1m, log\teff\,=\,3.843, log(L/L$_{\odot}$)\,=\,1.120), pulsation class ($\delta$\,Sct) and identified frequencies of UV\,PsA. \citet{2016AJ....152..207R} derived the atmospheric parameters and variable type of UV\,PsA using HARPS spectra as \teff\,=\,7918 K, \logg\,=\,4.00, [Fe/H]\,=\,0.18 and \ds\,. \citet{2022MNRAS.516.2080B} determined the dominate frequency as 9.15\,d$^{-1}$ for UV\,PsA in their catalog of bright 372 \ds\, stars and their period-luminosity relation using TESS and Gaia DR3 data.

\setcounter{table}{0}
\renewcommand{\thetable}{A\arabic{table}}
\begin{table*}
    \centering
    \caption{The result of the analysis of chemical abundances. The numbers given in the brackets represent the number of used lines in the analysis.}\label{tab:table_abundance}
    \begin{tabular}{llllll}
    \hline
Element & AU\,Scl & FG\,Eri & HZ Vel & MX Vir & IO Lup \\ 
\hline
$_{6}$C & 8.44\,$\pm$\,0.35 (2) & 8.81\,$\pm$\,0.29 (5) & 8.49\,$\pm$\,0.33 (6) & 8.50\,$\pm$\,0.35 (7)  &   \\ 
$_{11}$Na &  & 6.74\,$\pm$\,0.35 (1) & 6.07\,$\pm$\,0.26 (2) & 6.26\,$\pm$\,0.27 (3) &     \\ 
$_{12}$Mg& 7.58\,$\pm$\,0.31 (4) & 8.04\,$\pm$\,0.24 (3) & 7.25\,$\pm$\,0.28(6) & 7.73\,$\pm$\,0.20 (8) & 7.82\,$\pm$\,0.38 (2) \\ 
$_{14}$Si& 7.49\,$\pm$\,0.39 (14) & 7.10\,$\pm$\,0.33 (9) & 6.65\,$\pm$\,0.23 (9) & 7.16\,$\pm$\,0.53 (17) & 6.30\,$\pm$\,0.38  (2) \\ 
$_{16}$S&  &  &  & 7.37\,$\pm$\,0.35 (2) &  \\ 
$_{20}$Ca & 6.55\,$\pm$\,0.34 (6) & 6.47\,$\pm$\,0.31 (11) & 5.93\,$\pm$\,0.31 (20) & 6.35\,$\pm$\,0.17 (23) & 6.29\,$\pm$\,0.38 (2)  \\ 
$_{21}$Sc & 1.64\,$\pm$\,0.32 (2) & 3.91\,$\pm$\,0.28 (4) & 2.60\,$\pm$\,0.29 (6) & 3.46\,$\pm$\,0.23 (8) & 3.17\,$\pm$\,0.36 (3)  \\ 
$_{22}$Ti & 5.62\,$\pm$\,0.30 (8) & 5.30\,$\pm$\,0.31 (23) & 4.55± 0.22 (33) & 4.95\,$\pm$\,0.37 (83) & 5.10± 0.30 (24) \\ 
$_{23}$V&   & 5.00\,$\pm$\,0.30 (2) & 4.11\,$\pm$\,0.26 (7) & 4.09\,$\pm$\,0.32 (8) & 4.41\,$\pm$\,0.35 (3) \\ 
$_{24}$Cr & 6.34\,$\pm$\,0.39 (20) & 6.03\,$\pm$\,0.27 (14) & 5.18\,$\pm$\,0.24 (33) & 5.67\,$\pm$\,0.53 (90) & 5.52\,$\pm$\,0.35 (9) \\ 
$_{25}$Mn& 5.41\,$\pm$\,0.25 (5) & 5.54\,$\pm$\,0.23 (4) & 5.00\,$\pm$\,0.28 (7) & 5.39\,$\pm$\,0.27 (24) & \\
$_{26}$Fe& 8.01\,$\pm$\,0.17 (60) & 7.67\,$\pm$\,0.16 (49) & 6.95\,$\pm$\,0.12(104) & 7.43\,$\pm$\,0.12 (369) & 7.44\,$\pm$\,0.14 (56)  \\ 
$_{27}$Co& ~ & ~ & 5.01\,$\pm$\,0.33 (2) & 5.04\,$\pm$\,0.32 (2) & 6.12\,$\pm$\,0.38 (2)  \\ 
$_{28}$Ni& 6.98\,$\pm$\,0.29 (14) & 6.58\,$\pm$\,0.31 (15) & 5.82\,$\pm$\,0.29 (24) & 6.31\,$\pm$\,0.35 (84) & \\ 
$_{29}$Cu& ~ & 4.09\,$\pm$\,0.35 (2) & ~ & 3.99\,$\pm$\,0.35 (2) &  \\ 
$_{30}$Zn& ~ & ~ & ~ & 4.66\,$\pm$\,0.35 (2) & \\
$_{38}$Sr& ~ & 3.61\,$\pm$\,0.74 (1) & 3.55\,$\pm$\,0.68 (1) & 3.64\,$\pm$\,0.36 (2) & \\ 
$_{39}$Y& 4.18\,$\pm$\,0.34 (2) & 3.07\,$\pm$\,0.35 (2) & 2.20\,$\pm$\,0.37 (2) & 3.29\,$\pm$\,0.35 (6) & \\ 
$_{40}$Zr& ~ & 2.97\,$\pm$\,0.36 (4) & 2.51\,$\pm$\,0.23 (6) & 3.46\,$\pm$\,0.32 (7) & \\ 
$_{56}$Ba& 4.63\,$\pm$\,0.36 (2) & 3.00\,$\pm$\,0.35 (2) & 2.08\,$\pm$\,0.38 (2) & 3.32\,$\pm$\,0.35 (3) &  \\ 
\hline
Element & KU Com & V527 Car & V1187 Tau & UV PsA & DP UMa \\
\hline
$_{6}$C & ~ & 8.69\,$\pm$\,0.52 (3) & 8.27\,$\pm$\,0.29 (4)  & 8.58\,$\pm$\,0.32 (4) & 8.34\,$\pm$\,0.33 (6)  \\ 
$_{11}$Na & 6.21\,$\pm$\,0.52 (2) & ~ & 6.23\,$\pm$\,0.35 (3) & 5.65\,$\pm$\,0.28 (1) & 6.21\,$\pm$\,0.27 (3)  \\ 
$_{12}$Mg& 7.64\,$\pm$\,0.23 (4) & 7.59\,$\pm$\,0.53 (4) & 7.58\,$\pm$\,0.12 (4) & 7.92\,$\pm$\,0.36 (3) & 7.65\,$\pm$\,0.27 (6)  \\ 
$_{14}$Si& 7.28\,$\pm$\,0.48 (8) & 7.36\,$\pm$\,0.52 (9) & 7.15\,$\pm$\,0.25 (11) & 7.12\,$\pm$\,0.27 (13) & 7.05\,$\pm$\,0.25 (19) \\ 
$_{16}$S&  &  &  &  & 7.41\,$\pm$\,0.30 (3) \\ 
$_{20}$Ca & 6.38\,$\pm$\,0.50 (15) & 6.23\,$\pm$\,0.50 (10) & 6.31\,$\pm$\,0.26 (14) & 6.82\,$\pm$\,0.26 (13) & 5.82\,$\pm$\,0.27 (18) \\
$_{21}$Sc & 2.99\,$\pm$\,0.52 (5) & 2.94\,$\pm$\,0.52 (5) & 2.85\,$\pm$\,0.34 (5) & 3.63\,$\pm$\,0.25 (5) & 2.46\,$\pm$\,0.28 (6)\\ 
$_{22}$Ti & 4.86\,$\pm$\,0.53 (29) & 4.91\,$\pm$\,0.53 (29) & 4.74\,$\pm$\,0.18 (21) & 5.23\,$\pm$\,0.26 (17) & 4.78\,$\pm$\,0.29 (29) \\ 
$_{23}$V& 4.44\,$\pm$\,0.53 (3) & 4.54\,$\pm$\,0.53 (9) & 4.52\,$\pm$\,0.32 (2) & 4.75\,$\pm$\,0.35 (2) & 4.56\,$\pm$\,0.29 (4) \\
$_{24}$Cr & 5.43\,$\pm$\,0.55 (29) & 5.58\,$\pm$\,0.55 (25) & 5.52\,$\pm$\,0.30 (19) & 5.90\,$\pm$\,0.25 (17) & 5.69\,$\pm$\,0.26 (26)  \\ 
$_{25}$Mn& 4.87\,$\pm$\,0.54 (4) & 4.97\,$\pm$\,0.54 (9) & 5.34\,$\pm$\,0.25 (6) & 5.37\,$\pm$\,0.34 (7) & 5.04\,$\pm$\,0.22 (7)  \\
$_{26}$Fe& 7.29\,$\pm$\,0.14 (79) & 7.28\,$\pm$\,0.15 (91) & 7.30\,$\pm$\,0.15 (92) & 7.51\,$\pm$\,0.16 (61) & 7.31\,$\pm$\,0.15 (92)  \\ 
$_{27}$Co& ~ & ~ & 5.24\,$\pm$\,0.33 (1) & ~ & 5.71\,$\pm$\,0.35 (1)  \\ 
$_{28}$Ni& 6.09\,$\pm$\,0.27 (21) & 6.12\,$\pm$\,0.57 (18) & 6.08\,$\pm$\,0.31 (16) & 6.43\,$\pm$\,0.30 (25) & 6.23\,$\pm$\,0.25 (30) \\ 
$_{29}$Cu&  & 3.57\,$\pm$\,0.53 (1) & 3.26\,$\pm$\,0.34 (1) & 4.05\,$\pm$\,0.35 (2) & 4.00\,$\pm$\,0.25 (2) \\ 
$_{30}$Zn&  &  & ~ & 3.96\,$\pm$\,0.15 (1) & 4.45\,$\pm$\,0.25 (1) \\ 
$_{38}$Sr& ~ & 3.05\,$\pm$\,0.53 (1) & ~ & 3.79\,$\pm$\,0.26 (1) & \\ 
$_{39}$Y& ~ & ~ & 2.31\,$\pm$\,0.18 (4) & 3.12\,$\pm$\,0.62 (5) & 2.82\,$\pm$\,0.19 (5) \\
$_{40}$Zr& ~ & 2.46\,$\pm$\,0.53 (1) & 2.61\,$\pm$\,0.28 (2) & 4.39\,$\pm$\,0.64 (3) & 3.05\,$\pm$\,0.23 (3) \\ 
$_{56}$Ba& ~ & ~ & ~ & 3.54\,$\pm$\,0.25 (2) & 2.98\,$\pm$\,0.27 (1) \\ 
\hline
    \end{tabular}
\end{table*}

\setcounter{table}{1}
\begin{table*}
\centering
\caption{The pulsation frequencies of the targets. Formal error estimates are given in brackets in units of the last digits after the comma.}
\begin{tabular}{lccc|lccc}
\hline
\textbf{} & \multicolumn{3}{c|}{AU\,Scl} & \textbf{} & \multicolumn{3}{c}{FG\,Eri} \\
\cline{2-4} \cline{6-8}
ID & Frequency (d$^{-1}$) & Amplitude (mmag) & SNR & ID & Frequency (d$^{-1}$) & Amplitude (mmag) & SNR \\
\hline
$f_1$ & 16.5501 (1) & 1.66 & 46 & $f_1$ & 6.2921 (1) & 12.08 & 274 \\
$f_2$ & 19.3299 (1) & 1.49 & 42 & $f_2$ & 3.2367 (2) & 2.28 & 59 \\
$f_3$ & 15.7103 (1) & 1.11 & 36 & $f_3$ & 4.2721 (1) & 1.90 & 44 \\
$f_4$ & 12.7477 (2) & 0.90 & 46 & $f_4$ & 3.4718 (1) & 1.28 & 33 \\
$f_5$ & 12.8545 (2) & 0.79 & 42 & $f_5$ & 3.7962 (1) & 0.99 & 23 \\
$f_6$ & 18.1823 (2) & 0.75 & 25 & $f_6$ & 6.9964 (1) & 0.97 & 21 \\
$f_7$ & 14.9199 (3) & 0.55 & 27 & $f_7$ & 4.6368 (2) & 0.73 & 16 \\
$f_8$ & 18.5724 (4) & 0.37 & 12 & $f_8$ & 5.1924 (2) & 0.46 & 10 \\
$f_9$ & 16.5234 (3) & 0.41 & 11 & $f_9$ & 7.2574 (3) & 0.52 & 12 \\
$f_{10}$ & 13.7907 (6) & 0.26 & 16 & $f_{10}$ & 3.8729 (2) & 0.54 & 12 \\
$f_{11}$ & 17.0798 (5) & 0.28 & 8 & $f_{11}$ & 4.4477 (2) & 0.46 & 11 \\ 
$f_{12}$ & 16.3468 (6) & 0.25 & 7 & $f_{12}$ & 3.6887 (3) & 0.43 & 10 \\
$f_{13}$ & 17.5725 (7) & 0.23 & 7 & $f_{13}$ & 8.0318 (2) & 0.34 & 9 \\
$f_{14}$ & 16.0306 (6) & 0.23 & 7 & $f_{14}$ & 6.2786 (2) & 0.39 & 9 \\
$f_{15}$ & 19.2355 (6) & 0.23 & 6 & $f_{15}$ & 7.0732 (2) & 0.34 & 7 \\
$f_{16}$ & 17.7655 (6) & 0.22 & 6 & $f_{16}$ & 7.7698 (2) & 0.32 & 8 \\
$f_{17}$ & 15.1478 (7) & 0.21 & 8 & $f_{17}$ & 4.5341 (2) & 0.31 & 7 \\
$f_{18}$ & 16.4371 (9) & 0.21 & 6 & $f_{18}$ & 3.9113 (2) & 0.27 & 6 \\
$f_{19}$ & 22.0606 (9) & 0.17 & 5 & $f_{19}$ & 0.6986 (2) & 0.25 & 6 \\
$f_{20}$ & 17.7018 (9) & 0.19 & 5 & $f_{20}$ & 5.2624 (2) & 0.24 & 5 \\
\hline
\textbf{} & \multicolumn{3}{c|}{V1187\,Tau} & \textbf{} & \multicolumn{3}{c}{HZ\,Vel} \\
\cline{2-4} \cline{6-8}
ID & Frequency (d$^{-1}$) & Amplitude (mmag) & SNR & ID & Frequency (d$^{-1}$) & Amplitude (mmag) & SNR \\
\hline
$f_1$ & 53.1771 (1) & 1.61 & 249 & $f_1$ & 14.3398 (1) & 3.24 & 291 \\
$f_2$ & 46.1046 (2) & 1.42 & 225 & $f_2$ & 1.6155 (1) & 1.51 & 30 \\
$f_3$ & 49.6750 (2) & 0.81 & 138 & $f_3$ & 1.0665 (1) & 1.05 & 20 \\
$f_4$ & 46.1684 (2) & 0.33 & 50 & $f_4$ & 0.3542 (1) & 0.63 & 15 \\
$f_5$ & 46.3098 (2) & 0.29 & 44 & $f_5$ & 1.5977 (1) & 0.51 & 10 \\
$f_6$ & 49.8145 (1) & 0.24 & 41 & $f_6$ & 15.2400 (2) & 0.39 & 30 \\
$f_7$ & 46.5351 (1) & 0.22 & 34 & $f_7$ & 1.8104 (2) & 0.33 & 6 \\
$f_8$ & 0.0541 (1) & 0.22 & 15 & $f_8$ & 2.0003 (2) & 0.34 & 7 \\
$f_9$ & 53.1192 (1) & 0.18 & 28 & $f_9$ & 1.5285 (2) & 0.30 & 6 \\
$f_{10}$ & 41.3432 (1) & 0.14 & 30 & $f_{10}$ & 1.6640 (2) & 0.29 & 6 \\
$f_{11}$ & 41.1478 (1) & 0.12 & 26 & $f_{11}$ & 13.0191 (2) & 0.23 & 21 \\
$f_{12}$ & 37.5270 (1) & 0.12 & 24 & $f_{12}$ & 1.9796 (2) & 0.24 & 5 \\
$f_{13}$ & 43.4031 (1) & 0.12 & 22 & & & & \\
$f_{14}$ & 42.9542 (1) & 0.11 & 21 & & & & \\
$f_{15}$ & 0.0384 (1) & 0.10 & 6 & & & & \\
$f_{16}$ & 46.1130 (1) & 0.12 & 20 & & & & \\
$f_{17}$ & 49.4738 (1) & 0.09 & 16 & & & & \\
$f_{18}$ & 0.1342 (1) & 0.09 & 18 & & & & \\
$f_{19}$ & 42.3250 (1) & 0.08 & 5 & & & & \\
$f_{20}$ & 23.4808 (1) & 0.08 & 14 & & & & \\
$f_{21}$ & 0.1173 (1) & 0.07 & 15 & & & & \\
$f_{22}$ & 39.3151 (1) & 0.08 & 13 & & & & \\
$f_{23}$ & 42.6018 (1) & 0.07 & 15 & & & & \\
$f_{24}$ & 0.0691 (1) & 0.07 & 5 & & & & \\
\hline
\end{tabular}
\label{tab1:puls_freq}
\end{table*}

\setcounter{table}{1}
\begin{table*}
\centering
\caption{Continuation.}
\begin{tabular}{rccc|rccc}
\hline
& \multicolumn{3}{c|}{V527\,Car} & \multicolumn{3}{c}{DP\,UMa} \\
\cline{2-4} \cline{5-7}
& Frequency (d$^{-1}$) & Amplitude (mmag) & SNR & Frequency (d$^{-1}$) & Amplitude (mmag) & SNR \\
& $\pm 0.02$ & & & $\pm 0.02$ & & \\
\hline
$f_1$ & 4.6812 (1) & 15.05 & 207 & 18.0751 (1) & 2.03 & 49 \\
$f_2$ & 3.8243 (1) & 11.03 & 133 & 15.7013 (2) & 1.15 & 33 \\
$f_3$ & 3.5775 (1) & 3.13 & 41 & 17.4207 (1) & 0.98 & 26 \\
$f_4$ & 9.3603 (2) & 1.44 & 26 & 30.0962 (1) & 0.89 & 24 \\
$f_5$ & 0.1402 (2) & 1.27 & 12 & 28.4044 (1) & 0.88 & 22 \\
$f_1{-}f_2$ & 0.8569 (2) & 1.09 & 12 & 17.9254 (1) & 0.90 & 22 \\
$f_1{+}f_3$ & 8.2587 (3) & 1.02 & 14 & 18.7253 (2) & 0.82 & 16 \\
$2f_4{-}f_1{-}f_7$ & 5.7809 (4) & 1.03 & 17 & 2.5492 (2) & 0.82 & 16 \\
$2f_2$ & 7.6486 (3) & 0.83 & 11 & 18.9370 (3) & 0.70 & 17 \\
$f_{10}$ & 9.0642 (6) & 0.72 & 12 & 19.7218 (2) & 0.54 & 17 \\
$f_{11}$ & 0.2073 (5) & 0.58 & 6 & 12.0358 (2) & 0.63 & 12 \\
$f_{12}$ & 2.0099 (6) & 0.52 & 6 & $f_{10}{+}2f_1{-}2f_2$ & 24.4695 (3) & 0.63 & 16 \\
$f_{13}$ & 4.3771 (7) & 0.47 & 6 & 0.0342 (2) & 0.39 & 6 \\
$f_{14}$ & 1.1886 (6) & 0.50 & 5 & $2f_{13}{-}f_{10}{+}2f_2$ & 11.7493 (2) & 0.55 & 10 \\
$f_{15}$ & 12.3376 (6) & 0.43 & 7 & 18.8258 (2) & 0.52 & 12 \\
$f_{10}{-}f_{11}{-}f_6$ & 4.6535 (6) & 0.41 & 6 & 2.0295 (2) & 0.48 & 7 \\
$f_{17}$ & 4.8273 (7) & 0.39 & 6 & 16.5033 (2) & 0.43 & 12 \\
$f_{18}$ & 4.1935 (9) & 0.36 & 5 & 2.3246 (2) & 0.41 & 7 \\
& & & & $f_{15}{+}f_{18}{-}f_7$ & 2.4251 (2) & 0.44 & 8 \\
& & & & $f_{20}$ & 36.0029 (2) & 0.42 & 8 \\
& & & & $f_{21}$ & 2.8464 (2) & 0.38 & 8 \\
& & & & $f_{22}$ & 11.4798 (2) & 0.32 & 6 \\
& & & & $f_{23}$ & 1.5890 (2) & 0.32 & 5 \\
& & & & $f_{24}$ & 10.5624 (2) & 0.32 & 5 \\
\hline
& \multicolumn{3}{c|}{KU\,Com} & \multicolumn{3}{c}{MX\,Vir} \\
\cline{2-4} \cline{5-7}
& Frequency (d$^{-1}$) & Amplitude (mmag) & SNR & Frequency (d$^{-1}$) & Amplitude (mmag) & SNR \\
& $\pm 0.02$ & & & $\pm 0.02$ & & \\
\hline
$f_1$ & 4.3339 (1) & 0.36 & 16 & 6.4950 (1) & 14.15 & 318 \\
$f_2$ & 1.5763 (2) & 0.32 & 10 & 6.5949 (1) & 1.30 & 30 \\
$f_3$ & 2.2930 (2) & 0.23 & 9 & 5.9161 (1) & 1.16 & 26 \\
$f_4$ & 1.8980 (2) & 0.18 & 6 & 6.3117 (1) & 1.51 & 33 \\
$f_5$ & 1.7381 (2) & 0.20 & 7 & 12.9900 (1) & 0.99 & 35 \\
$f_6$ & 1.9995 (1) & 0.20 & 7 & 6.3575 (2) & 0.68 & 14 \\
$f_7$ & 2.6353 (1) & 0.15 & 6 & 6.2680 (2) & 0.63 & 13 \\
$f_8$ & 4.0217 (1) & 0.13 & 6 & 5.7661 (2) & 0.55 & 9 \\
$f_9$ & 19.9108 (1) & 0.14 & 15 & $f_7{+}2f_8{-}2f_3$ & 5.9681 (2) & 0.39 & 5 \\
$f_{10}$ & 2.1425 (1) & 0.17 & 7 & 5.8245 (2) & 0.36 & 8 \\
$f_{11}$ & 2.0861 (1) & 0.15 & 5 & $f_5{+}f_6{-}2f_4$ & 6.7240 (2) & 0.21 & 5 \\
$f_{12}$ & 4.2173 (1) & 0.11 & 5 & & & & \\
\hline
\end{tabular}
\label{tab1:puls_freq_1}
\end{table*}

\setcounter{table}{1}
\begin{table*}
\centering
\caption{Continuation.}
\begin{tabular}{rccc|rccc}
\hline
& \multicolumn{3}{c|}{IO\,Lup} & \multicolumn{3}{c}{UV\,PsA} \\
\hline
& Frequency (d$^{-1}$) & Amplitude (mmag) & SNR & Frequency (d$^{-1}$) & Amplitude (mmag) & SNR \\
& $\pm 0.02$ & & & $\pm 0.02$ & & \\
\hline
$f_1$ & 15.5946 (1) & 7.14 & 94 & 9.1511 (1) & 7.06 & 53 \\
$f_2$ & 15.4881 (1) & 3.91 & 49 & 8.3837 (2) & 5.06 & 40 \\
$f_3$ & 13.5090 (1) & 3.46 & 37 & 8.5227 (1) & 4.74 & 37 \\
$f_2 + f_3 - f_1$ & 13.4025 (2) & 2.04 & 22 & $f_4$ = 12.9809 (1) & 1.81 & 21 \\
$f_5$ & 15.1976 (2) & 1.73 & 20 & 13.5153 (1) & 1.43 & 19 \\
$f_6$ & 13.1255 (2) & 1.55 & 16 & 14.5789 (1) & 1.30 & 16 \\
$f_7$ & 15.2712 (3) & 1.57 & 18 & 17.0111 (2) & 1.22 & 14 \\
$f_8$ & 12.5291 (4) & 1.32 & 18 & 9.0969 (2) & 1.20 & 9 \\
$f_9$ & 15.0194 (3) & 1.30 & 15 & 7.4104 (3) & 1.11 & 14 \\
$f_2 + f_5 - f_9$ & 15.6662 (6) & 1.33 & 18 & $f_{10}$ = 7.8275 (2) & 0.88 & 10 \\
$f_{11}$ & 13.8304 (5) & 0.98 & 10 & 13.2752 (2) & 0.81 & 9 \\
$f_{12}$ & 12.7634 (6) & 0.94 & 11 & $f_3 + 2f_1 - 2f_4$ = 0.8631 (3) & 0.80 & 8 \\
$f_{13}$ & 15.6236 (7) & 0.69 & 8 & 16.2202 (2) & 0.79 & 9 \\
$f_{14}$ & 14.9304 (6) & 0.59 & 7 & 7.5458 (2) & 0.79 & 9 \\
$f_{15}$ & 13.5845 (6) & 0.54 & 5 & 8.1020 (2) & 0.66 & 6 \\
 & & & & $f_{14} + f_2 - f_{10}$ = 9.2901 (2) & 0.70 & 5 \\
 & & & & $f_1 + f_3 - f_2$ = 8.8062 (2) & 0.62 & 5 \\
\hline
\end{tabular}
\label{tab1:puls_freq_2}
\end{table*}

\end{document}